\def\eeq{\relax}
\def\beq#1#2\eeq{\begin{equation}\label{#1}#2\end{equation}}
\def\bal#1#2\eal{\begin{align}\label{#1}#2\end{align}}
\def\bse#1#2\ese{\begin{subequations}\label{#1}#2\end{subequations}}
\def\ba{\begin{aligned}}
\def\ea{\end{aligned}}

\def\dd{\operatorname{d}}

\newcommand{\al}{\alpha}

\newcommand{\de}{\delta}

\newcommand{\om}{\omega}

\newcommand{\si}{\sigma}

\newcommand{\pa}{\partial}

\newcommand{\been}{\begin{equation}}
\newcommand{\een}{\end{equation}}
\newcommand{\beena}{\begin{eqnarray}}
\newcommand{\eena}{\end{eqnarray}}

\documentclass[11pt]{article}

\usepackage{amssymb,mathtools}
\usepackage{graphicx}
\usepackage{subfigure}

\usepackage{color}

\def\C{\mathcal{C}}

\renewcommand{\appendix}{
  \setcounter{section}{0}\renewcommand{\thesection}{\Alph{section}}
  \section*{Appendix} }

\begin{document} 
\def\singlespacing{\baselineskip=13pt}	\def\doublespacing{\baselineskip=18pt}
\singlespacing

\pagestyle{myheadings}\markright{{\sc Active elastodynamic cloaking}  ~~~~~~\today}

\title{Active elastodynamic cloaking}

\author{Andrew N.\ Norris$^\dagger$, Feruza A.\ Amirkulova$^\dagger$, William J.\ Parnell$^{\dagger\dagger}$\footnote{Corresponding author: William.Parnell@manchester.ac.uk}
\\ \\
$\dagger$Mechanical and Aerospace Engineering, Rutgers University,\\
Piscataway, NJ 08854-8058, USA. \\
$\dagger\dagger$ School of Mathematics, Alan Turing Building, University of Manchester,\\ Oxford Road, Manchester, M13 9PL, UK.
}

\maketitle

 \begin{abstract}

An active elastodynamic cloak destructively interferes with an incident time harmonic in-plane (coupled compressional/shear) elastic wave to produce zero total elastic field over a finite spatial region. A method is described which  explicitly predicts the source amplitudes of the active field. For a given number of sources and their positions in two dimensions it is shown that the multipole amplitudes can be expressed as infinite sums of the coefficients of the incident wave decomposed into regular Bessel functions. Importantly, the active field generated by the sources vanishes in the far-field. In practice the infinite summations are clearly required to be truncated and  the accuracy of cloaking is studied when the truncation parameter is modified.

\end{abstract}


\section{Introduction} \label{sec1}

The main function of a cloaking device is to render an object invisible to some incident wave as seen by some external observer. Over the past decade, a great deal of effort has been focused on passive cloaking, using metamaterials to guide waves around specific regions of space, see e.g.\ the highly cited works  \cite{Leonhardt06,Pendry06,Cummer07}. In recent times a rather different approach to cloaking has been noted as an alternative. It has been named {\it active exterior cloaking} and it relies on a set of discrete active sources, lying outside the cloaking region, to nullify the incident wave whilst their own radiated field must be negligible in the far-field. Interest has focused on the Helmholtz equation in two dimensions \cite{Miller06,Vasquez09,Vasquez09b,Vasquez11a,Vasquez11,Norris12b}. In the work of Vasquez et al.\  \cite{Vasquez09,Vasquez09b} Green's formula and addition theorems for Bessel functions were used to formulate an integral equation, which was then converted to a linear system of equations for the unknown amplitudes.  Crucially, the integral equation provides the source amplitudes as linear functions of the incident wave field.  It was shown that active cloaking can be realized using as few as three active sources in 2D. Further work to render the linear relation for the source amplitudes in more explicit form was developed in \cite{Vasquez11} and extended to the three dimensional Helmholtz case in  \cite{Vasquez11a}. In \cite{Norris12b}, the integral representations of Vasquez et al.\  \cite{Vasquez11} for  the source amplitudes were reduced to closed-form explicit formulas. This obviated the need to reduce the integral equation of Vasquez et al.\  \cite{Vasquez09,Vasquez09b} to a system of linear equations which are then required to be solved numerically or to evaluate line integrals, as proposed in  \cite{Vasquez11}.

There is, of course, a strong link between active exterior cloaking and the notion of \textit{anti-sound} or in the context of elastic media \textit{anti-vibration}.  Interestingly the notion of anti-sound appears to have been considered first in a patent published in 1936 by Paul Lueg, see e.g.\ \cite{Gui-90}. The subject has focused greatly on the desire to reduce the magnitude of a radiating field or to create so-called quiet zones in enclosed domains such as aircraft cabins using simple sources.  The idea to suppress completely the sound field in a finite volume inside an unbounded domain using the Kirchhoff-Helmholtz integral formula and thus employing a continuous distribution of monopoles and dipoles is described in \cite{Nel-92}. Anti-vibration techniques have also been developed \cite{McK-89,Ful-96}. In general the focus of anti-sound is to reduce the sound radiated from a sound source or to create a zone of silence by employing a finite number of radiating sources. The active field is not required to be non-radiating however. Furthermore very little work in the anti-sound community has focused on the exact shape of the quiet zone with the exception of \cite{Dav-94} who calculated, numerically the zone of silence ($< 10$dB) region created when the amplitude of a single secondary source was chosen to reduce noise due to a single primary source.

The aim of active exterior cloaking is to render the total field zero inside some prescribed domain (the \textit{cloak} or \textit{zone of silence}), whilst ensuring that the active field itself is non-radiating. The technique introduced in the early active exterior cloaking work enables a cloaked region to be identified clearly by the use of Graf's addition theorem. This approach allows precise determination of the necessary source amplitudes.

The infinite series associated with the multipole expansion of the $m$th active source is formally divergent inside the circle that is centered on source itself, i.e.\ for $|{\bf x}-{\bf x}_m|< a_m$ in the notation used later on. Therefore the representation for the source is not valid in the domain in which it resides! This point has not been stressed in the active cloaking community, although a related point was noted in the anti-sound community in \cite{Kem-76}. Clearly this point motivates the truncation of the series which has to be done practically in any case. This limits accuracy but as we shall see later in many cases, only a small number of multipoles are required.

As yet it does not appear that active exterior cloaking has been applied to the elastodynamic context. This paper will focus on the relevant two dimensional active elastodynamic cloaking problem. In general, elastodynamic cloaking problems are more difficult to study than their acoustic or electromagnetic counterparts. Indeed in the case of passive elastodynamic cloaking, this is due to the lack of invariance of Navier's equations under coordinate transformations \cite{Milton06} unless we relax the minor symmetry property of the required elastic modulus tensor. The latter can be achieved by using Cosserat materials \cite{Brun09,Norris11a} or by employing nonlinear pre-stress of hyperelastic materials \cite{Parnell2011,Parnell2012,Norris2012aa}. Here we show how the active approach to cloaking can be employed in the elastodynamic case for the fully coupled two-dimensional (in-plane) compressional/shear wave problem. As in the approach of \cite{Norris12b} we write down the relevant integral equation by employing the isotropic Green's tensor in this case. The required source amplitudes for arbitrary wave incidence can be determined explicitly by using Graf's addition theorem.

We shall begin in \S\ref{sec2} with a statement of the problem, a review of the governing  equations, and a summary of the main results.  The relevant integral relation is derived in \S\ref{sec3}, from which the main results regarding the explicit form of the source amplitudes are shown to follow.  We consider both compressional and transverse (shear) wave incidence. We also describe the form of the active source field and the issues associated with divergence described above. Numerical results follow in \S\ref{sec4}.

\section{Problem formulation and main results}  \label{sec2} 

\subsection{Problem overview}

\begin{figure}[htb]	
	\centering
		\includegraphics[width=4.5in]{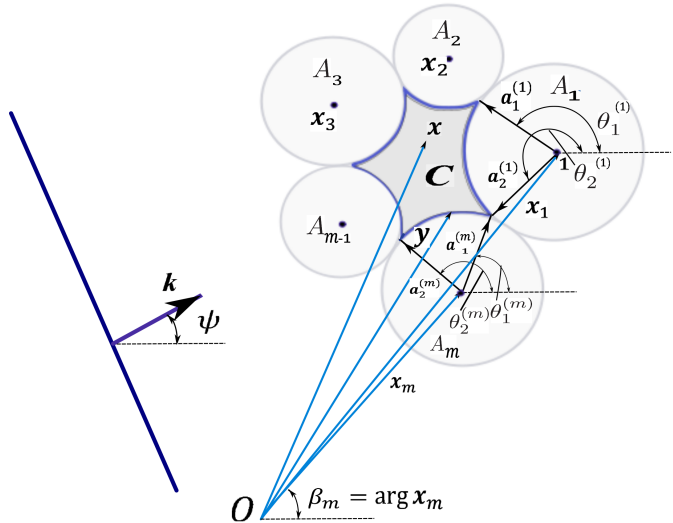}
\caption{Insonification of the  actively cloaked region $\C$ generated by  $M$ active point multipole sources  at ${\bf x}_m$,
and {active sources regions $\mathcal{A}_m$,  $m=\overline{1,M}$.}
The incident field in this case is a  plane wave with wave vector $\textbf{k}$
in the direction  $\psi$.}
\label{fig:fig_1}
\end{figure}

Let us consider the two-dimensional configuration where the active cloaking devices consist of arrays of point multipole sources located at positions ${\bf x}_m \in \mathbb{R}^2$, $m=\overline{1,M}$ as depicted in
 Figure \ref{fig:fig_1}. These sources can give rise to both shear and compressional elastic waves. The active sources lie in the exterior region with respect to the  cloaked region $\C$
and for this reason, this type of cloaking is called \textit{active exterior cloaking} \cite{Vasquez09}.
Objects are undetectable in the cloaked region by virtue of the destructive interference of the sources and  the incident field with the result that  the total wave amplitude  vanishes in the cloaked region $\C$. As described in \cite{Norris12b} this gives rise to three significant advantages over passive cloaking: (i) the cloaked region is not completely surrounded by a single cloaking device; (ii) only a small  number of active sources are needed; (iii)  the procedure  works for broadband input sources.  The principal disadvantage of the  method is of course that  the incident field must be known.

The $M$ active sources give rise to a cloaked zone $\C$ is indicated in Figure \ref{fig:fig_1} by the shaded region whose boundary $\C$ is the closed concave union of the circular arcs  $\partial C_m$ $m=\overline{1,M}$, $\{ a_m, \theta^{(m)}_{1}, \theta^{(m)}_{2} \}$  associated with the source at ${\bf x}_m$. In the general case $\{ a_m, \theta^{(m)}_{1}, \theta^{(m)}_{2} \}$ are distinct for different values of $m$.  Note that the wave incidence shown in Figure \ref{fig:fig_1} is a plane wave  although the solution  derived below is for arbitrary incidence. We therefore have to determine the amplitudes of the active sources as a function of the incident wave, and then prove that the cloaked region is indeed the closed region $\C$ as indicated in Figure \ref{fig:fig_1}. Let us also define the notation $\mathcal{A}_m$ as the circular domain of radius $a_m$ that contains the $m$th active source at its centre. We also define the union of these domains $\mathcal{A}=\cup_{m=1}^M \mathcal{A}_m$.

\subsection{P/SV in-plane wave propagation}

We consider time harmonic solutions with the factor $e^{-i\omega t}$ understood but omitted.
Navier's equations in two dimensions for the displacement ${\bf u} = (u_1,u_2)$,
$u_j = u_j(x_1,x_2)$,  are,
\beq{-11}
\partial_j \sigma_{ij} + \rho \omega^2 u_i = f_i,
\quad
\sigma_{ij} = C_{ijkl} \, \partial_l u_k,
\eeq
where $\mathbf{f}$ is the forcing,
$C_{ijkl} = \lambda \delta_{ij}\delta_{kl} + \mu (\delta_{ik}\delta_{jl} +\delta_{il}\delta_{jk} )$ and the summation convention on repeated indices is understood.
Hence, in the absence of  forcing ($\mathbf{f} =0$),
\beq{406}
(\lambda+2\mu)\nabla \nabla\cdot {\bf u}+\mu\nabla^2 {\bf u} + \rho\om^2  {\bf u} = 0.
\eeq
 The Helmholtz decomposition for the displacement,
\beq{222}
\mathbf{u} = \nabla\Phi + \nabla\!\!\times\!(\Psi\mathbf{k})
\eeq
leads to separate Helmholtz equations for the scalar potentials
\begin{align}
\nabla^2\Phi + k_p^2\Phi = 0, \qquad
\nabla^2\Psi + k_s^2\Psi = 0
\end{align}
where $k_p$, $k_s$ are the longitudinal and shear wavenumbers, respectively: $k_p^2 = \omega^2\rho/(\lambda+2\mu)$, $k_s^2=\omega^2\rho/\mu$.
We also  define, for later use,
$\kappa \equiv  k_s/k_p$, or equivalently
$\kappa^2 = 2(1-\nu)/(1-2\nu)$ where $\nu $ is Poisson's ratio.
We seek the total wave field in the form of an incident wave, ${\bf u}_i$, plus the active source  field,
${\bf u}_d$, such that
\begin{align}
{\bf u} ({\bf x} ) = {\bf u}_i + {\bf u}_d
\quad
\Rightarrow
\quad
\Phi ({\bf x} )= \Phi_i + \Phi_d,
\qquad
\Psi ({\bf x} ) = \Psi_i + \Psi_d .
\end{align}
We assume the general form of an incident field in the regular basis, and hence
 \bse{49}
 \bal{-49a}
\begin{pmatrix}
\Phi_i
\\
\Psi_i
\end{pmatrix}
=& \sum_{n=-\infty}^{\infty}
 \begin{pmatrix}
 A_n^{(p)} U_n^{\,+}(k_p {\bf x} )  	
 \\
 A_n^{(s)} U_n^{\,+}(k_s {\bf x} ) 	
\end{pmatrix} ,
\\
\begin{pmatrix}
\Phi_d
\\
\Psi_d
\end{pmatrix}
=&
\sum\limits_{m=1}^M\sum\limits_{n=-\infty}^\infty
\begin{pmatrix}
 B_{m,n}^{(p)} V_n^{\,+}(k_p ({\bf x}- {\bf x}_m) )  		
 \\
 B_{m,n}^{(s)} V_n^{\,+}(k_s ({\bf x}- {\bf x}_m) )  		
 \end{pmatrix}, \label{-49b}
 \eal
 \ese
  where  the functions $U_n^{\,\pm}({\bf z} )$  and   $V_n^{\,\pm}({\bf z} )$  are defined by
\beq{12-}
U_n^{\,\pm}({\bf z} ) = J_n (|{\bf z}|) e^{\pm i n \arg {\bf z} },
\qquad
V_n^{\,\pm}({\bf z} ) = H_n^{(1)} (|{\bf z}|) e^{\pm i n \arg {\bf z} }.
\eeq
Here $\arg {\bf z} \in [ 0, 2 \pi)$ and $\arg {(- \,\bf z)} = \arg {\bf z} \pm \pi  \in [ 0, 2 \pi)$.
Define the derivative functions ${U_n^{\,\pm}}'({\bf z})$ and ${V_n^{\,\pm}}'({\bf z})$ as
\beq{8-}
{U_n^{\,\pm}}'({\bf z} ) = J_n'(|{\bf z}|)e^{\pm i n \arg {\bf z} }, \quad \quad {V_n^{\,\pm}}'({\bf z} ) = {H_n^{(1)}}'(|{\bf z}|)e^{\pm i n \arg {\bf z} }.
\eeq
Note that the  functions
 $U_n^{\,\pm}({\bf z} )$  and   $V_n^{\,\pm}({\bf z} )$   possess the properties
\beq{13-}
U_n^{\,\pm}(-  {\bf z} ) = (-1)^n U_n^{\,\pm}({\bf z} ),
\qquad
V_n^{\,\pm}(- {\bf z} ) = (-1)^n V_n^{\,\pm}({\bf z} ).
\eeq
In the following we write $U_0$ and $V_0$, with obvious meaning.

\subsection{Summary of the main results} 			\label{Summary}

Here we shall state the main results and the required source amplitudes to enable perfect active cloaking together with necessary and sufficient conditions on these amplitudes. The latter ensures we can compare accuracy of the cloaking technique. We shall prove these results in \S\ref{sec3}. Let $\{ a_m, \theta^{(m)}_{1}, \theta^{(m)}_{2} \}$ define the circular arc   $\partial C_m$ of the closed boundary of the cloaked region associated with the source at ${\bf x}_m$. 
The active source amplitude coefficients
for the general form of an incident field \eqref{-49a}  are
\begin{subequations}\label{-9}
\bal{-9a}
\begin{pmatrix}
B_{m,\,l}^{(p)}
\\ \\
B_{m,\,l}^{(s)}
\end{pmatrix}
=
&
\sum\limits_{n=-\infty}^\infty
\begin{pmatrix}
B_{m,\,ln}^{(p)} \,A_{n}^{(p)}
\\ \\
B_{m,\,ln}^{(s)} \,A_{n}^{(s)}
\end{pmatrix},
 \quad \text{where}
\\
\begin{pmatrix}
B_{m,\,ln}^{(p)}
\\ \\
B_{m,\,ln}^{(s)}
\end{pmatrix}
=
& \frac{1}{4   {(k_s a_m)}^2  }  \sum\limits_{q=-\infty}^\infty    (-1)^q
\big[ e^{-i (q+l) \theta^{(m)}_{2} } - e^{-i (q+l) \theta^{(m)}_{1} } \big]
\notag \\
&    \cdot
\bigg\{
 U_{n+q}^{\,+}(k_p {\bf x}_m)
\begin{pmatrix}
v_1 (k_p a_m, \, k_s a_m)
\\
v_2 (k_p a_m, \, k_s a_m)
\end{pmatrix}
+
U_{n+q}^{\,+}(k_s {\bf x}_m)
\begin{pmatrix}
-v_2 (k_s a_m, \, k_p a_m)
\\
\,\,\,\,\, v_1 (k_s a_m, \, k_p a_m)
\end{pmatrix}
\bigg\}, \label{-9b}
\\
{\bf v} (\alpha , \beta)
=&
\begin{pmatrix}
v_1
\\
v_2
\end{pmatrix}=
\begin{pmatrix}
  \big[\frac{{\alpha_s^{(m)}}^2}{q+l} - 2q \big]  { \alpha } J_l'(\alpha ) &
i \big[\frac{{\alpha_s^{(m)}}^2}{q+l} - 2l \big] {\alpha}  J_l(\alpha )
\\ & \\
 -i  \big[{\alpha_s^{(m)}}^2  - 2lq\big]    {J_l}(\beta) &
 - 2   \alpha_p^{(m)} \alpha_s^{(m)} {J_l}'(\beta)
\end{pmatrix}
\begin{pmatrix}
J_q (\alpha )
\\ \\
i  J_q' (\alpha )
\end{pmatrix} .     \label{-9c}
\eal
\end{subequations}
Derivation of equation \eqref{-9} is given in Section \ref{superposition}. Alternatively defining a vector {${\bf a}_i^{(m)} \equiv  a_m \hat{\bf e}(\theta_i^{(m)}) (i=1,\,2)$,}  and incorporating equations \eqref{12-} and \eqref{8-}, equation \eqref{-9b} reduces to the form
\bse{+9=1}
\bal{+9}
\begin{pmatrix}
B_{m,\,ln}^{(p)}
\\ \\
B_{m,\,ln}^{(s)}
\end{pmatrix}
=
& \frac{1}{4 {\alpha_s^{(m)}}^2 }  \sum\limits_{q=-\infty}^\infty
  (-1)^q \bigg\{
U_{n+q}^{\,+}(k_p {\bf x}_m)
\begin{pmatrix}
V_1 (k_p {\bf a}, \, k_s {\bf a})
\\
V_2 (k_p {\bf a}, \, k_s {\bf a})
\end{pmatrix}														\nonumber
\\
& \qquad \qquad  \qquad \qquad
+
U_{n+q}^{\,+}(k_s {\bf x}_m)
\begin{pmatrix}
-V_2 (k_s {\bf a}, \, k_p {\bf a})
\\
\,\,\,\,\,V_1 (k_s {\bf a}, \, k_p {\bf a})
\end{pmatrix}
\bigg\} \bigg|_{{\bf a}_1^{(m)}}^{{\bf a}_2^{(m)} },  \quad \text{where}                    
\\
{\bf V} (\boldsymbol{\alpha} , \boldsymbol{\beta})  =
&
\begin{pmatrix}
V_1
\\
V_2
\end{pmatrix}=
\begin{pmatrix}
  \big[\frac{{\alpha_s^{(m)}}^2}{q+l} - 2q \big]  { {\alpha} } {U_l^{-}}' ({ \boldsymbol{\alpha} }) &
i \big[\frac{{\alpha_s^{(m)}}^2}{q+l} - 2l \big]  { {\alpha} }  {U_l^{-}}({ \boldsymbol{\alpha} } )
\\ & \\
 -i  \big[{\alpha_s^{(m)}}^2  - 2lq\big]    {U_l^{-}}(\boldsymbol{\beta}) &
 - 2   \alpha_p^{(m)} \alpha_s^{(m)} {U_l^{-}}'(\boldsymbol{\beta})
\end{pmatrix}
\begin{pmatrix}
{U_q^{-}} ({ \boldsymbol{\alpha} } )
\\ \\
i {U_q^{-}}' ({ \boldsymbol{\alpha} } )
\end{pmatrix} .					\label{+9b}
\eal
\ese

Next, we note that the active source coefficients $B_{m,\,l}^{(p)}$ and $B_{m,\,l}^{(s)}$ must satisfy the necessary and sufficient conditions to ensure active cloaking inside the domain $\C$.
\beq{7}
	\forall \, n\in \mathbb{Z}: \qquad
\sum\limits_{m=1}^M\sum\limits_{l=-\infty}^\infty
\,
\times
\begin{cases}
  B_{m,\,l}^{(p)} U_{n-l}^{\,-}({k_p\bf x}_m) &= 0,
	\\ &  \\
	B_{m,\,l}^{(s)} U_{n-l}^{\,-}(k_s{\bf x}_m) &= 0,
	  \\ &\\
	  B_{m,\,l}^{(p)}  V_{n-l}^{-}(k_p{\bf x}_m)  &= -A_n^{(p)},
	  \\&\\
	  B_{m,\,l}^{(s)}	   V_{n-l}^{-}(k_s{\bf x}_m)  &= -A_n^{(s)}.
\end{cases}	
\eeq
The first pair of conditions are required to ensure zero radiated field outside the union of the active regions, $\mathcal{A}$, and the second pair ensures that the total field is zero inside $\C$.
These constraints on  $B_{m,\,l}^{(p)}$ and $B_{m,\,l}^{(s)}$  will be  used to estimate the error in the active cloaking region in the following sections by truncating the infinite sums in \eqref{7}.


\section{Derivation of the source amplitude expressions and constraints}     \label{sec3}  

Let us first formulate the problem in terms of an integral equation.

\subsection{{Green's} tensor and integral equation formulation}

Consider the  particular solution of Navier's equations \eqref{-11} in the presence of a point force,
\beq{uGF}
\mathbf{f} = \mathbf{F} \de({\bf x}-{\bf x}_0)
\quad \Rightarrow \quad
u_i = G_{ik}F_k
\eeq
where $G_{ik}$ is the two dimensional (in-plane) Greens tensor.
Specifically, $G_{ik}({\bf x}) $ satisfies
\beq{4=5}
\Sigma_{ijk,j} + \rho \omega^2 G_{ik} = \delta_{ik} \delta({\bf x}),
\quad
\Sigma_{ijk} = C_{ijpq}G_{pk,q},
\eeq
with solution
\beq{4=44}
G_{ik} = 
\big(\rho \omega^2\big)^{-1}\, \Big[
\de_{ik} \, k_s^2G_s + \partial_i \partial_k (G_s-G_p)
\Big]
\eeq
where
\begin{align}
G_s = \frac{1}{4i}V_0(k_s {\bf x}),
\qquad
G_p = \frac{1}{4i}V_0(k_p {\bf x}).
\end{align}
The solution \eqref{4=44} can be checked by substitution into the governing equation
\eqref{4=5} and using the identities
$(\nabla^2 + k_\alpha^2)G_\alpha = \delta({\bf x}),$ $\alpha = p,s$.

It is convenient to work without subscripts, writing \eqref{4=44} as
\beq{1-2}
-\rho \omega^2
{\bf G}( {\bf x}) = \nabla \nabla G_p + ({\bf I}\nabla^2 - \nabla \nabla) G_s
= \nabla \nabla G_p + ( \nabla \!\!\times\! {\bf k})( \nabla \!\!\times\! {\bf k}) G_s
\ \ \text{for}\   {\bf x} \ne 0 ,
\eeq
where $( \nabla \!\!\times\! {\bf k})_i = e_{ij3}\partial_j$.  Using \eqref{uGF}
in the form ${\bf u} = {\bf G}\cdot {\bf F}$, combined with
the Helmholtz decomposition for $\mathbf{u}$ gives
\beq{111}
 \nabla\Phi + (\nabla\!\!\times\! \mathbf{k}) \Psi
  = 
	\big(\rho \omega^2\big)^{-1}\,
  \big[ \nabla \nabla G_p \cdot {\bf F} +
  ( \nabla \!\!\times\! {\bf k})( \nabla \!\!\times\! {\bf k}) G_s \cdot {\bf F}
  \big],
  \ \   {\bf x} \ne 0 ,
 \eeq
implying
\beq{5=2=1}
\begin{aligned}
-\rho \omega^2 \Phi &= {\bf F}\cdot   \nabla G_p
= F_1\partial_1 {G_p}  + F_2 \partial_2 {G_p},
\\
-\rho \omega^2\Psi &={\bf F}\cdot   ( \nabla \!\!\times\! {\bf k}) G_s
=F_1\partial_2 {G_s}  - F_2 \partial_1 {G_s} ,
\end{aligned}
\ \qquad   {\bf x} \ne 0 .
\eeq
This makes it clear that for a standard point source, regardless of the choice of $\mathbf{F}$,
 both compressional and shear waves propagate away from the point source.

With knowledge of the Green's tensor we can now develop an integral equation for the displacement. Indeed, if ${\bf u} $ is a solution of the homogeneous equations in an infinite domain containing a finite region $D$ and
${\boldsymbol \sigma}$ is the associated stress, then by definition of the Green's tensor,
\beq{5=6}
\int_{\partial D} \dd S n_i \big[
u_j({\bf y}) \Sigma_{ijk}({\bf y}-{\bf x} )
- \sigma_{ij}({\bf y}) G_{jk}({\bf y}-{\bf x} )
\big]
= \begin{cases}
u_k({\bf x}), & \quad {\bf x} \in D , \\
0, & \quad {\bf x} \notin D .
\end{cases}
\eeq
Equation \eqref{5=6} holds for both ${\bf u}_i$ and ${\bf u}_d$ separately inside the cloaked region, since both are assumed to be regular there (this is a definition of exterior cloaking).
Also, by its definition   the total field is zero inside the cloaked region with boundary $\partial C$, and therefore
\beq{5=8}
{\bf u}_d ({\bf x}) = -
\int_{\partial C} \dd S {\bf n} \cdot \big[
{\bf u}_i ({\bf y}) \cdot {\boldsymbol \Sigma}({\bf y}-{\bf x} )
- {\boldsymbol \sigma}_i ({\bf y})  \cdot {\bf G}({\bf y}-{\bf x} )
\big],  \quad {\bf x} \in C.
\eeq
This is the fundamental relation used to find the source amplitudes.

\subsection{General expressions for the source amplitudes}

Following the procedure for the Helmholtz problem \cite{Norris12b}, we first substitute the assumed form of ${\bf u}_d$ into the left member of
\eqref{5=8}.  Then we partition  the integral in the right member into $M$ segments over
$\{ {\partial C}_m , m=\overline{1,M} \}$ and identify each line integral with the
$m^\text{th}$ component of ${\bf u}_d $, i.e.\ the part of the source field from the multipoles at ${\bf x}_m$.  Thus,
\bal{5=9}
& 0=
\sum_{m=1}^M \bigg\{
\int_{\partial C_m} \dd S {\bf n} \cdot \big[
{\bf u}_i ({\bf y})\cdot {\bf \Sigma}({\bf y}-{\mathbf x} )
- {\bf \sigma}_i ({\bf y}) \cdot {\bf G}({\bf y}-{\bf x} )
\big]
\notag \\
&\quad +\sum\limits_{n=-\infty}^\infty
\bigg(
 B_{m,n}^{(p)} \nabla V_n^{\,+}(k_p ({\bf x}- {\bf x}_m) )
 +
 B_{m,n}^{(s)}\nabla \!\!\times\! {\bf k} V_n^{\,+}(k_s ({\bf x}- {\bf x}_m) )
 \bigg)
 \bigg\}
,  \quad {\bf x} \in C.
\eal

We now use the generalized Graf addition theorem \cite[eq.\ (9.1.79)]{Abramowitz74},
\beq{-2}
V_l^{\,+}({\bf y} - {\bf x})
= \sum\limits_{n=-\infty}^\infty
\begin{cases}
V_n^{\,+}({\bf y} )\,
 U_{n-l}^{-}({\bf x}) ,   & |{\bf y}|>|{\bf x}|,
\\
U_n^{\,+}({\bf y} )\,
 V_{n-l}^{-}({\bf x})   , & |{\bf y}|<|{\bf x}|.
\end{cases}
\eeq
The idea is to write ${\bf \Sigma}({\bf y}-{\bf x} )$ and ${\bf G}({\bf y}-{\bf x} ) $ in \eqref{5=9}
in terms of sources at ${\bf x}_m$.  This suggests
using \eqref{-2} for  ${\bf y}-{\bf x} = ({\bf y}-{\bf x}_m) - ({\bf x}-{\bf x}_m)$
subject to $|{\bf y}-{\bf x}_m|<   |{\bf x}-{\bf x}_m|$.
Hence, using \eqref{1-2},
\bal{1-3}
{\bf G}({\bf y}-{\bf x} ) =& \frac i{4\rho \omega^2}
\sum\limits_{n=-\infty}^\infty
\bigg\{
\nabla \nabla
U_n^{-}\big( k_p({\bf y}-{\bf x}_m ) \big)
 V_{n}^{\,+}\big( k_p({\bf x} -{\bf x}_m) \big)
 \notag \\  &  \qquad \qquad
+ ( \nabla \!\!\times\! {\bf k})( \nabla \!\!\times\! {\bf k})
U_n^{-}\big( k_s({\bf y}-{\bf x}_m ) \big)
 V_{n}^{\,+}\big( k_s({\bf x} -{\bf x}_m) \big)
\bigg\}.
\eal
By virtue of the dependence of the Green's function on ${\bf y} -{\bf x}$ the derivatives
$\nabla \nabla $ can be understood as $\nabla_y \nabla_y $ or $\nabla_x \nabla_x $
or $-\nabla_y \nabla_x $, with the same equivalence for
$( \nabla \!\!\times\! {\bf k})( \nabla \!\!\times\! {\bf k})$.    Inspection of \eqref{5=9} suggests that the forms $-\nabla_y \nabla_x $   and
$-( \nabla_y \!\!\times\! {\bf k})( \nabla_x \!\!\times\! {\bf k})$ are appropriate.
   Taking into account the negative sign in
 $\nabla \nabla \to -\nabla_y \nabla_x $,  the Green's function can be written in the form
\bal{1-4}
{\bf G}({\bf y}-{\bf x} ) =& \frac {-i}{4\rho \omega^2}
\sum\limits_{n=-\infty}^\infty
\bigg\{
\nabla_y
U_n^{-}\big( k_p({\bf y}-{\bf x}_m ) \big)
 \nabla_x V_{n}^{\,+}\big( k_p({\bf x} -{\bf x}_m) \big)
 \notag
 \\  &  \qquad \qquad
+ ( \nabla_y \!\!\times\! {\bf k})
U_n^{-}\big( k_s({\bf y}-{\bf x}_m ) \big) ( \nabla_x \!\!\times\! {\bf k})
 V_{n}^{\,+}\big( k_s({\bf x} -{\bf x}_m) \big)
\bigg\}.
\eal

Substituting from \eqref{1-4}  into \eqref{5=9},  and
 identifying the coefficients of
 $\nabla V_n^{\,+}(k_p ({\bf x}- {\bf x}_m) )$
and
$\nabla \!\!\times\! {\bf k} V_n^{\,+}(k_s ({\bf x}- {\bf x}_m) )  $, yields
\bse{422}
\bal{4-2}
 B_{m,n}^{(p)} &=
 \frac {-i}{4\rho \omega^2}
 \int_{\partial C_m} \!\!\dd S {\bf n} \cdot \big[
 {\boldsymbol \sigma}_i ({\bf y}) \cdot
\nabla
U_n^{-}\big( k_p({\bf y}-{\bf x}_m ) \big)
- {\bf u}_i ({\bf y})\cdot
{\bf \sigma}^{(p)} \big( k_p({\bf y}-{\bf x}_m ) \big)
\big],
\\
B_{m,n}^{(s)} &= \frac {-i}{4\rho \omega^2}
 \int_{\partial C_m} \!\! \dd S {\bf n} \cdot \big[
 {\boldsymbol \sigma}_i ({\bf y}) \cdot
( \nabla \!\!\times\! {\bf k})
U_n^{-}\big( k_s({\bf y}-{\bf x}_m ) \big)
- {\bf u}_i ({\bf y})\cdot
{\bf \sigma}^{(s)} \big( k_s({\bf y}-{\bf x}_m ) \big)
\big] ,
\eal
where
\beq{9=9}
\ba
 \sigma^{(p)}_{ij} \big( k_p({\bf y}-{\bf x}_m ) \big)
&= C_{ijpq} U_{n,pq}^{-}\big( k_p({\bf y}-{\bf x}_m ) \big) ,
\\
 \sigma^{(s)}_{ij} \big( k_s({\bf y}-{\bf x}_m ) \big)
&= C_{ijpq} e_{pr3}
 U_{n,rq}^{-}\big( k_s({\bf y}-{\bf x}_m ) \big) .
\ea
\eeq
\ese

Therefore, given the incident field, we are now able to evaluate the required source amplitudes that guarantee zero total field inside the domain $\C$.  We can however, make further progress on the integrals in \eqref{422} in order to render them in simpler form, by using the fact that $\partial C_m$ is the arc of the circle of radius $a_m$ centered at ${\bf x}_m$, which is the origin of the shifted coordinates
${\bf y}-{\bf x}_m $.  The integration is therefore simplified using polar coordinates centered at  ${\bf x}_m$, combined with the expressions for
 the displacements and traction components in polar coordinates
 given in terms of the potentials,
\beq{390}
\ba
u_r &= \Phi_{,r} + \frac 1r \Psi_{,\theta},
\quad
u_\theta = \frac 1r \Phi_{,\theta} - \Psi_{,r},
\\
\si_{rr} &= -\lambda k_p^2\Phi + 2\mu\big(\Phi_{,rr}+\frac{1}{r}\Psi_{,r\theta}-\frac{1}{r^2}\Psi_{,\theta}\big),
\\
\si_{r\theta} &= 2\mu\big(\frac{1}{r}\Phi_{,r\theta}-\frac{1}{r^2}\Phi_{,\theta}\big)
+ \mu\big(\frac{1}{r^2}\Psi_{,\theta\theta}-\Psi_{,rr}+\frac{1}{r}\Psi_{,r}\big).
\ea
\eeq
The four distinct terms in the integrals of \eqref{422}, such as
$\dd S {\bf n} \cdot    {\boldsymbol \sigma}_i ({\bf y}) \cdot
 \nabla  U_n^{-}\big( k_p({\bf y}-{\bf x}_m ) \big)$,
then follow by identifying $\Phi \to U_n^{-}\big( k_p{\bf a} )$, $\Psi \to U_n^{-}\big( k_s{\bf a})$,
where ${\bf a} (\theta) \equiv {\bf y}-{\bf x}_m$ is the radial vector of constant magnitude $a_m$.
 Thus,
\bse{422=11}
\bal{4=21}
\dd S {\bf n} \cdot
 {\boldsymbol \sigma}_i   \cdot
 \nabla
U_n^{-}
&=\dd S \Big[  {\sigma}_{irr} \frac {\partial}{\partial r} U_n^{-}\big( k_p{\bf a} \big) +  {\sigma}_{ir \theta} \frac{1}{r} \frac {\partial}{\partial \theta}
U_n^{-}\big( k_p{\bf a} \big)\Big]
\notag \\
&=
\dd \theta\, \Big(\sigma_{irr} k_p a_m {U_n^{-}}'(k_p {\bf a}) -  in    \sigma_{ir\theta} U_n^{-} (k_p {\bf a})\Big),
 \displaybreak[0]
\\		
\dd S {\bf n}  \cdot
{\boldsymbol \sigma}^{(p)} \cdot {\bf u}_i
&=\dd S \big[ u_{ir} {\sigma}_{rr}^{(p)}+ u_{i \theta} {\sigma}_{r \theta}^{(p)} \big]
\notag \\
& = \dd \theta  \frac{\mu}{a_m}\Big(
 u_{ir}  \Big[ \big( 2n^2-k_s^2 a_m^2  \big)  {U_n^{-}}(k_p {\bf a})
- 2 k_p a_m{{U_n}^{-}}'(k_p {\bf a})\Big]
\notag \\
& \qquad \qquad
+
 u_{i\theta}\,  2 i n    \big[ {U_n^{-}}(k_p {\bf a}) - k_p a_m {{U_n}^{-}}'(k_p {\bf a}) \big]
\Big), 
 \displaybreak[0]
\\  		
\dd S {\bf n} \cdot {\boldsymbol \sigma}_i  \cdot
( \nabla \!\!\times\! {\bf k})
U_n^{-}
&
=\dd S \Big[ {\sigma}_{irr}\frac{1}{r} \frac {\partial}{\partial \theta}  U_n^{-}\big( k_s{\bf a} \big) - {\sigma}_{ir\theta} \frac {\partial}{\partial r}
U_n^{-}\big( k_s{\bf a} \big)\Big]
\notag \\
& =- \dd \theta\Big(in \sigma_{irr}  {U_n^{-}}(k_s {\bf a})+ k_s a_m \sigma_{ir\theta} {U_n^{-}}'(k_s {\bf a}) \Big),
 \displaybreak[0]
\\		
\dd S {\bf n}  \cdot
{\boldsymbol \sigma}^{(s)}    \cdot {\bf u}_i
 &=\dd S \big[ u_{ir} {\sigma}_{rr}^{(s)}+ u_{i \theta} {\sigma}_{r \theta}^{(s)} \big]
\notag
\\
&
=\dd \theta  \frac{\mu}{a_m}\Big( u_{ir}\,2in \big[ {U_n^{-}}(k_s {\bf a}) - k_s a_m {U_n^{-}}'(k_s {\bf a}) \big] \notag
\\
& \quad \quad + u_{i\theta} \big[ \big( (k_s a_m)^2 - 2 n^2 \big){U_n^{-}}(k_s {\bf a})+ 2 k_s a_m {U_n^{-}}'(k_s {\bf a})\big) \Big) .
\eal
\ese

Noting the reversal of the sense of the integral in equation \eqref{422} and incorporating equation \eqref{4=21} leads to
\bse{4=22}
 \bal{-4=22a}
B_{m,\,l}^{(p)} =& \frac{1} {4k_s^2}\int_{\theta_1^{(m)}}^{\theta_2^{(m)}} \!\!\dd \theta \, e^{-i\,l\,\theta}
\bigg\{
 i\alpha_p^{(m)}{J_l}'(\alpha_p^{(m)}) \frac{\sigma_{irr}}{\mu}  +\,l\,  {J_l}(\alpha_p^{(m)})   \frac{\sigma_{ir\theta}}{\mu}	 \nonumber
\\
&
+ i \Big[ \big( {\alpha_s^{(m)}}^2 - 2\, l^2 \big)
 {J_l}(\alpha_p^{(m)})  + 2 \alpha_p^{(m)}  {J_l}'(\alpha_p^{(m)}) \Big] \frac{u_{ir}}{a_m}
+  2\, l
\Big[  {J_l}(\alpha_p^{(m)}) -   \alpha_p^{(m)} {J_l}'(\alpha_p^{(m)}) \Big] \frac{u_{i\theta}}{a_m}
\bigg\},
\\
B_{m,\,l}^{(s)} =& \frac{1} {4k_s^2}\int_{\theta_1^{(m)}}^{\theta_2^{(m)}} \!\!\dd \theta \, e^{-i\,l\,\theta}
\bigg\{
-i \alpha_s^{(m)}{J_l}'(\alpha_s^{(m)})  \frac{\sigma_{ir\theta}}{\mu}
 + l {J_l}(\alpha_s^{(m)}) \frac{\sigma_{irr}}{\mu}
\nonumber
\\
&
-
i \Big[  \big( {\alpha_s^{(m)}}^2-2\, l^2 \big){J_l}(\alpha_s^{(m)})  + 2\,\alpha_s^{(m)} {J_l}'(\alpha_s^{(m)}) \Big]
 \frac{u_{i\theta}}{a_m}
+ 2\, l\, \Big[  {J_l}(\alpha_s^{(m)}) - \alpha_s^{(m)} {J_l}'(\alpha_s^{(m)})\Big] \frac{u_{ir}}{a_m}
\bigg\}, 								\label{-4=22b}
\eal
 \ese
where $\alpha_p^{(m)}=k_p \,a_m$,
$\alpha_s^{(m)}=k_s\, a_m$,
$\theta_1^{(m)}$ and
$\theta_2^{(m)}$ are the angular positions of the  vectors
${\bf a}_i^{(m)} \equiv a_m {\hat{ \bf e}}(\theta_i^{(m)})$, $i=1,2$,
which describe the  initial and final positions of segment $\partial C_m$.
Equations \eqref{4=22} provide expressions for the source amplitudes for any time harmonic incident field.

Let us now specialize the result to the specific case of plane wave incidence. This is important in its own right but also allows us to derive the general incident wave case  by integration as we shall show.

\subsection{Plane wave incidence}

Let us define
\beq{0-2}
u_{\psi_{\al}}({\bf x}) = e^{ik_{\al} \hat{\mathbf{e}}(\psi_{\al})\cdot{\bf x}}, \quad  \al=p,s
\eeq
where $\hat{\mathbf{e}}(\psi_{\al})=(\cos{\psi_{\al}},\sin{\psi_{\al}})$ so that $u_{\psi_{\al}}$ correspond to compressional (p) and shear (s) plane waves of unit amplitude.

\subsubsection{Longitudinal incident plane wave}
Consider now longitudinal plane wave incidence
\beq{4=23}
\Phi_i  ({ \bf x})  = A_p u_{{\psi}_{p}}({\bf x})
\eeq
where $ A_p\equiv const $ is a known wave amplitude.
Then using the relation $\Phi_i ({ \bf y}) =\Phi_i ({\bf x}_m) u_{\psi_p}({\bf a}) $
with ${\bf a}= a_m {\hat{\bf e}}(\theta)$, and eq.\ \eqref{390} with $\Phi = \Phi_i $, $\Psi =0$,
  reduces equation \eqref{4=22}    to the form:
\bse{4=24}
 \bal{-4=24a}
B_{m,\,l}^{(p)} &= \frac{  \Phi_i({\bf x}_m)}{4\kappa  \alpha_s^{(m)} } \int_{\theta_1^{(m)}}^{\theta_2^{(m)}} \!\!\dd \theta \, e^{-i\,l\,\theta} u_{\psi_p}({\bf a})
 \bigg\{ i {\alpha_p^{(m)}}^2 {J_l}'(\alpha_p^{(m)}) \big[ 2  \sin^2(\theta-\psi_p) -\kappa^2 \big]  \nonumber
\\
&\quad \quad +  l \alpha_p^{(m)} {J_l}(\alpha_p^{(m)}) \sin 2(\theta-\psi_p)
- i 2 l \sin(\theta-\psi_p) \Big[  {J_l}(\alpha_p^{(m)})  - \alpha_p^{(m)}{J_l}'(\alpha_p^{(m)})\Big] \nonumber
\\
&\quad \quad  -\cos(\theta-\psi_p) \Big[ \big( {\alpha_s^{(m)}}^2-2\, l^2\big) {J_l}(\alpha_p^{(m)})
       + 2  \alpha_p^{(m)} {J_l}'(\alpha_p^{(m)}) \Big]
\bigg\},
\\
B_{m,\,l}^{(s)} &= \frac{ \Phi_i({\bf x}_m)}{4\kappa  \alpha_s^{(m)} }
\int_{\theta_1^{(m)}}^{\theta_2^{(m)}} \!\!\dd \theta \, e^{-i\,l\,\theta} u_{\psi_p}({\bf a})
\cdot\bigg\{  \big[2 \sin^2(\theta-\psi_p)-\kappa^2 \big] \, l \, \alpha_p^{(m)}  {J_l}(\alpha_s^{(m)})\nonumber
\\
& \quad \quad - i\sin 2(\theta-\psi_p) \alpha_p^{(m)}  \alpha_s^{(m)} {J_l}'(\alpha_s^{(m)})
+ i2\, l\,\cos(\theta-\psi_p)\bigg[    {J_l}(\alpha_s^{(m)})  -  \alpha_s^{(m)} {J_l}'(\alpha_s^{(m)})	 \bigg]								 \nonumber
\\
&\quad \quad - \sin(\theta-\psi_p) \Big[\big( {\alpha_s^{(m)}}^2-2\, l^2\big) {J_l}(\alpha_s^{(m)})
     + 2 \alpha_s^{(m)} {J_l}'(\alpha_s^{(m)}) \Big] \bigg\}.	\label{-4=24b}
\eal
 \ese
Then noting that $u_{\psi_p}({\bf a})  = e^{ik_p a_m \cos (\theta - \psi_p)}
 = e^{i \alpha_p^{(m)} \cos (\theta - \psi_p)}$,
equation \eqref{4=24} can be written
\bse{4=25}
 \bal{-4=25a}
B_{m,\,l}^{(p)} &= \frac{i\Phi_i({\bf x}_m)}{4\kappa \alpha_s^{(m)} }
 e^{-i\,l\,\psi_p}
\cdot\bigg\{ {\alpha_p^{(m)}}^2 {J_l}'(\alpha_p^{(m)}) \Big[2 L_0''(\alpha_p^{(m)})  - \big(\kappa^{2}-2\big) L_0(\alpha_p^{(m)})  \Big]  \nonumber
\\
&\quad \quad- 2\,l \alpha_p^{(m)} {J_l}(\alpha_p^{(m)})  L_1'(\alpha_p^{(m)})
-   2 l L_1(\alpha_p^{(m)}) \Big[  {J_l}(\alpha_p^{(m)}) - \alpha_p^{(m)} {J_l}'(\alpha_p^{(m)}) \Big]
\nonumber
\\
&\quad \quad
+   {L_0}'(\alpha_p^{(m)})
\Big[\big( {\alpha_s^{(m)}}^2-2\, l^2\big) {J_l}(\alpha_p^{(m)})   + 2  \alpha_p^{(m)} {J_l}'(\alpha_p^{(m)}) \Big]\bigg\},
\\
B_{m,\,l}^{(s)} &= \frac{\Phi_i({\bf x}_m)}{4\kappa \alpha_s^{(m)} } e^{-i\,l\,\psi_p}
\cdot\bigg\{ l \alpha_p^{(m)} {J_l}(\alpha_s^{(m)})\Big[2 {L_0}''(\alpha_p^{(m)})    - \big(\kappa^{2}-2\big) L_0(\alpha_p^{(m)})  \Big]  \nonumber
\\
& \quad \quad -  2\,\alpha_p^{(m)} \alpha_s^{(m)}  {J_l}'(\alpha_s^{(m)})  L_1'(\alpha_p^{(m)})
+2\,l\, L_0'(\alpha_p^{(m)}) \Big[  {J_l}(\alpha_s^{(m)})- \alpha_s^{(m)} {J_l}'(\alpha_s^{(m)}) \Big]					 
		\nonumber
\\
&\quad \quad
-L_1(\alpha_p^{(m)})
\Big[ \big( {\alpha_s^{(m)}}^2-2\, l^2  \big) {J_l}(\alpha_s^{(m)})   + 2 \alpha_s^{(m)}  {J_l}'(\alpha_s^{(m)}) \Big]\bigg\},	\label{-4=25b}
\eal
 \ese
where the functions $L_0(\alpha)$ and $L_1(\alpha)$ are defined by
\beq{4=26}
L_j(\alpha) =
\int_{\theta_1^{(m)}-\psi_p }^{\theta_2^{(m)}-\psi_p } \dd \theta \,
( \sin \theta)^j\,
e^{i ( \alpha \cos \theta-   l \theta)}  ,
\quad j=0,1.
\eeq
 $L_0(\alpha)$  can be evaluated by using  the Jacobi-Anger identity $e^{i x \sin\theta} = \sum_{n=-\infty}^\infty J_n(x) e^{i n\theta}$,
\bal{403}
L_0(\alpha)
&=
 \sum\limits_{n=-\infty}^\infty
J_n(\alpha) \, i^{n}
\int_{\theta_1^{(m)}-\psi_p }^{\theta_2^{(m)}-\psi_p} \dd \theta \,
e^{-i ( n+   l) \theta }
\notag \\
&=
\sum\limits_{n=-\infty}^\infty
J_n(\alpha) \,  i^{n+1}\, \frac{ e^{i ( n+l)\psi_p } }{n+l} \big[ e^{-i (n+l) \theta^{(m)}_{2} } - e^{-i (n+l) \theta^{(m)}_{1} } \big] .
\eal
Integration by parts yields $L_1(\alpha)$ in the form
\beq{489}
L_1(\alpha) = - \frac{l}{\alpha} L_0(\alpha)  - \frac 1{i\alpha}
\left. e^{i ( \alpha \cos \theta-   l \theta )}
\right|_{\theta_1^{(m)}-\psi_p }^{\theta_2^{(m)}-\psi_p}.
\eeq
Taking into account the Jacobi-Anger identity and equation \eqref{403}, the function $L_1(\alpha)$ and its derivative $L_1'(\alpha)$ can be expressed
\bse{404}
 \bal{404a}
L_1(\alpha)
&= \frac{1}{\alpha}
 \sum\limits_{n=-\infty}^\infty
J_n(\alpha) \,  n\,i^{n+1}\,  \frac{ e^{i ( n+l)\psi_p } }{n+l}\cdot\big[ e^{-i (n+l) \theta^{(m)}_{2} } - e^{-i (n+l) \theta^{(m)}_{1} } \big] ,
\\
L_1'(\alpha)
&=  \frac{1 }{\alpha^2}
 \sum\limits_{n=-\infty}^\infty  n\,i^{n+1}
\big[ \alpha{J_n}'(\alpha)- {J_n}(\alpha)\big]\, \frac{ e^{i ( n+l)\psi_p } }{n+l}\cdot\big[ e^{-i (n+l) \theta^{(m)}_{2} } - e^{-i (n+l) \theta^{(m)}_{1} } \big].
\eal
 \ese
Introducing the explicit results for  the functions $L_0(\alpha)$ and $L_1(\alpha)$ into \eqref{4=25} yields  expressions for the amplitude coefficients in the form:
\bse{4=27}
 \bal{-4=27a}
B_{m,\,l}^{(p)} &= \frac{\Phi_i({\bf x}_m)}{4\kappa^{2} }  \sum\limits_{q=-\infty}^\infty   \frac{i^{q+2}e^{i   q  \psi_p} }{q+l}
\cdot \bigg\{ \alpha_p^{(m)} {J_l}'(\alpha_p^{(m)})  \bigg[ 2 {J_q}''(\alpha_p^{(m)})   - \big( {\kappa}^2 - 2 \big) J_q(\alpha_p^{(m)})   \bigg]  \nonumber
\\
&\qquad \qquad - \frac{2\,l\,q}{\alpha_p^{(m)}}  {J_l}(\alpha_p^{(m)})\bigg[{J_q}'(\alpha_p^{(m)}) -  \frac{1}{\alpha_p^{(m)}}  J_q(\alpha_p^{(m)}) \bigg]
+ J_q'(\alpha_p^{(m)}) \bigg[  2 {J_l}'(\alpha_p^{(m)})      \nonumber
\\
&\qquad \qquad
+\frac{{\alpha_s^{(m)}}^2-2 l^2}{\alpha_p^{(m)}}  {J_l}(\alpha_p^{(m)}) \bigg]-  \frac{2l\,q} {\alpha_p^{(m)}}
J_q(\alpha_p^{(m)})
 \Big[  \frac{1}{\alpha_p^{(m)}}{J_l}(\alpha_p^{(m)}) - {J_l}'(\alpha_p^{(m)}) \Big]\bigg\}  \notag
 \\
&\qquad \qquad
\cdot\big[ e^{-i (q+l) \theta^{(m)}_{2} } - e^{-i (q+l) \theta^{(m)}_{1} } \big] ,
\\  
B_{m,\,l}^{(s)} &= \frac{\Phi_i({\bf x}_m)}{4 \kappa } \sum\limits_{q=-\infty}^\infty   \frac{i^{q+1} e^{i   q  \psi_p} }{q+l}
\cdot \bigg\{\frac{ l}{\kappa} {J_l}(\alpha_s^{(m)})\bigg[ 2 {J_q}''(\alpha_p^{(m)})  -\big( {\kappa}^2 - 2 \big) J_q(\alpha_p^{(m)}) \bigg]\nonumber
\\
&
 \qquad \qquad
 - 2 \,q\,{J_l}'(\alpha_s^{(m)}) \bigg[{J_q}'(\alpha_p^{(m)})- \frac{1}{\alpha_p^{(m)}}  J_q(\alpha_p^{(m)}) \bigg]
+  2l {J_q}'(\alpha_p^{(m)}) \bigg[\frac{1}{\alpha_s^{(m)}}    {J_l}(\alpha_s^{(m)})  \notag
\\
&
 \qquad \qquad  - {J_l}'(\alpha_s^{(m)}) \bigg]  -    \frac{q}{\alpha_p^{(m)}}  J_q(\alpha_p^{(m)})
 \Big[ \frac{{\alpha_s^{(m)}}^2-2\, l^2 }{\alpha_s^{(m)}}{J_l}(\alpha_s^{(m)})   + 2   {J_l}'(\alpha_s^{(m)}) \Big]
\bigg\}     \notag
\\
&
 \qquad \qquad
\cdot\big[ e^{-i (q+l) \theta^{(m)}_{2} } - e^{-i (q+l) \theta^{(m)}_{1} } \big] .	\label{-4=27b}
\eal
 \ese
After some simplification eq.\ \eqref{4=27} can be written as
\bal{-23}
\begin{pmatrix}
B_{m,\,l}^{(p)}
\\ \\
B_{m,\,l}^{(s)}
\end{pmatrix}
=& \frac{\Phi_i({\bf x}_m)}{4   {\alpha_s^{(m)}}^2  }  \sum\limits_{q=-\infty}^\infty
 {i^{q} e^{i   q  \psi_p}}
\cdot\big[ e^{-i (q+l) \theta^{(m)}_{2} } - e^{-i (q+l) \theta^{(m)}_{1} } \big]
\nonumber
\\
& \ \  \cdot
\begin{pmatrix}
  \big[\frac{{\alpha_s^{(m)}}^2}{q+l} - 2q \big]  { \alpha_p^{(m)} J_l'(\alpha_p^{(m)})} &
i \big[\frac{{\alpha_s^{(m)}}^2}{q+l} - 2l \big]  { \alpha_p^{(m)}} J_l(\alpha_p^{(m)})
\\  & \\
- i  \big[{\alpha_s^{(m)}}^2  - 2lq\big]    {J_l}(\alpha_s^{(m)}) &
- 2   \alpha_p^{(m)}\alpha_s^{(m)} {J_l}'(\alpha_s^{(m)})
\end{pmatrix}
\begin{pmatrix}
J_q(\alpha_p^{(m)})
\\ \\
i J_q' (\alpha_p^{(m)})
\end{pmatrix}.
\eal

\subsubsection{ {Transverse} plane wave incidence}
Consider now an incident transverse plane wave
\beq{4=30}
\Psi_i= A_s e^{ik_s {\hat {\bf e}}(\psi_s)\cdot\bf{x}},
\eeq
where $ A_s\equiv const $ is a known transverse wave amplitude. Entirely analogous calculations to the compressional wave case yield the source amplitudes in the form
\bal{-22}
\begin{pmatrix}
B_{m,\,l}^{(s)}
\\ \\
-B_{m,\,l}^{(p)}
\end{pmatrix}
& = \frac{\Psi_i({\bf x}_m)}{4   {\alpha_s^{(m)}}^2  }  \sum\limits_{q=-\infty}^\infty
 {i^{q} e^{i   q  \psi_s}}
\big[ e^{-i (q+l) \theta^{(m)}_{2} } - e^{-i (q+l) \theta^{(m)}_{1} } \big]
\nonumber
\\
&  \ \  \cdot
\begin{pmatrix}
  \big[\frac{{\alpha_s^{(m)}}^2}{q+l} - 2q \big]  { \alpha_s^{(m)} J_l'(\alpha_s^{(m)})} &
i \big[\frac{{\alpha_s^{(m)}}^2}{q+l} - 2l \big] {\alpha_s^{(m)}}  J_l(\alpha_s^{(m)})
\\ & \\
 -i  \big[{\alpha_s^{(m)}}^2  - 2lq\big]    {J_l}(\alpha_p^{(m)}) &
 - 2   \alpha_p^{(m)} \alpha_s^{(m)} {J_l}'(\alpha_p^{(m)})
\end{pmatrix}
\begin{pmatrix}
J_q (\alpha_s^{(m)})
\\ \\
i J_q' (\alpha_s^{(m)})
\end{pmatrix}.
\eal

\subsubsection{Plane wave incidence summarized}

Adding the separate  results of  eqs.\ \eqref{-23} and \eqref{-22} gives for combined incidence
 \beq{463}
\Phi_i=A_p e^{ik_p {\bf \hat e}(\psi_p)\cdot\bf{x}},
\quad
\Psi_i=A_s e^{ik_s {\bf \hat e}(\psi_s)\cdot\bf{x}},
\eeq
the source amplitudes
\bal{-25}
\begin{pmatrix}
B_{m,\,l}^{(p)}
\\ \\
B_{m,\,l}^{(s)}
\end{pmatrix}
=& \frac{1}{4   {\alpha_s^{(m)}}^2  }  \sum\limits_{q=-\infty}^\infty
 i^{q}
\big[ e^{-i (q+l) \theta^{(m)}_{2} } - e^{-i (q+l) \theta^{(m)}_{1} } \big]
\notag \\
&    \cdot
\bigg\{
\Phi_i({\bf x}_m)e^{i   q  \psi_p}
\begin{pmatrix}
v_1 (\alpha_p^{(m)}, \alpha_s^{(m)})
\\
v_2 (\alpha_p^{(m)}, \alpha_s^{(m)})
\end{pmatrix}
+
\Psi_i({\bf x}_m)e^{i   q  \psi_s}
\begin{pmatrix}
-v_2 (\alpha_s^{(m)}, \alpha_p^{(m)})
\\
v_1 (\alpha_s^{(m)}, \alpha_p^{(m)})
\end{pmatrix}
\bigg\}
\eal
where the vector ${\bf v} (\alpha , \beta)  = (v_1,\, v_2)^T$ is defined in \eqref{-9c}.

\subsection{Arbitrary incident field as superposition of plane incident waves } \label{superposition}

The general form of incident field given by equation \eqref{-49a} can be constructed as superposition of plane incident waves of the form \eqref{463}.
This will enable us to find the general form of the amplitude coefficients for incident waves of general form as a superposition of solutions for plane waves given by \eqref{-25}. Recall the  incident field for a combined incident plane wave having the form:
\beq{49X}
\begin{pmatrix}
\Phi_i({\bf x})
\\
\Psi_i({\bf x})
\end{pmatrix}
=
\begin{pmatrix}
A_p e^{ik_p \hat{\bf e}(\psi_p)\cdot {\bf x}}
\\
A_s e^{ik_s \hat{\bf e}(\psi_s)\cdot {\bf x}}
\end{pmatrix}
=\sum_{q =-\infty}^{\infty}
 \begin{pmatrix}
 i^q e^{-iq \psi_p} U_q^{\,+}(k_p {\bf x} )  	
 \\
 i^q e^{-iq \psi_s} U_q^{\,+}(k_s {\bf x} ) 	
\end{pmatrix} .
 \eeq

Multiplying the first row of equation \eqref{49X} by $(i^{-(n+q)}/2\pi) e^{i(n+q) \psi_p}$ and the second row by $(i^{-(n+q)}/2\pi) e^{i(n+q) \psi_s}$, integrating with respect to $\psi_p$ and $\psi_s$ respectively between $0$ and $2\pi$ and then evaluating at ${\bf x} ={\bf x}_m$ we find
\bse{49X1}
\bal{-49X1a}
 \frac{i^{-(n+q)}} {2\pi}\int_0^{2\pi} \dd \psi_p \, \Phi_i({\bf x}_m) e^{i(n+q)\psi_p}
&= U_{n+q}^{\,+}( {k_p {\bf x}_m} ),
\\
\frac{i^{-(n+q)}} {2\pi}\int_0^{2\pi} \dd \psi_s \, \Psi_i({\bf x}_m) e^{i(n+q)\psi_s}
 &=
  U_{n+q}^{\,+}( {k_s{ \bf x}_m} ).
 \eal
 \ese
To obtain the form of the amplitude coefficients given by eq. \eqref{-9} for the general incidence \eqref{-49a} we multiply the first and second of equations \eqref{-25} by $A_n^{(p)}\gamma_n(\psi_p,\psi_s)$ and $A_n^{(s)}\gamma_n(\psi_p,\psi_s)$ respectively, where $\gamma_n(\psi_p,\psi_s)=i^{-2n}/(2\pi)^2e^{in\psi_p}e^{in\psi_s}$, carry out the double integration with respect to $\psi_p$ and $\psi_s$ between $0$ and $2\pi$, incorporate \eqref{49X1} and sum over all $n\in\mathbb{Z}$.

\subsection{Necessary and sufficient conditions on the source amplitudes}

In this section we will define the constraints on the active source coefficients $B_{m,n}^{(p)}$ and $B_{m,n}^{(s)}$ by  expressing the active source field ${ \bf u}_d$ in terms of near-field and far-field source amplitudes and using Graf theorem \eqref{-2}.  When $|{\bf x}|>|{\bf y}|$ the components of ${\bf u}_d$ can be defined as a sum of multipoles at the origin using the first identity in \eqref{-2}
\beq{3}
\left.
\ba
\Phi_d &=\sum\limits_{n=-\infty}^\infty F_{n}^{(p)} V_n^{\,+}(k_p{\bf x}),
\\
\Psi_d &= \sum\limits_{n=-\infty}^\infty F_{n}^{(s)} V_n^{\,+}(k_s{\bf x}),
\ea
\right\}
\quad
\ \ \text{ for }
\ \  |{\bf x}|> \text{max}(|{\bf x}_m|+a_m),
\eeq
where
 \bal{-4a}
F_{n}^{(p)} =\sum\limits_{m=1}^M  \sum\limits_{l=-\infty}^\infty B_{m,\,l}^{(p)} U_{n-l}^{\,-}({k_p\bf x}_m),
\quad
F_{n}^{(s)} =\sum\limits_{m=1}^M  \sum\limits_{l=-\infty}^\infty B_{m,\,l}^{(s)} U_{n-l}^{\,-}(k_s{\bf x}_m). 
 \eal
If the active field $\Phi_d$ and $\Psi_d$ does not radiate into the far-field, then we must have $F_{n}^{(p)}=0, \,\, F_{n}^{(s)}=0,\, \forall \, n$ ensuring the necessity of \eqref{7}$_{1,2}$. Sufficiency is guaranteed by substituting the expressions \eqref{7}$_{1,2}$ into an assumed far-field of the form \eqref{3}.

Next we consider the near-field. Assuming $|{\bf x}_m|>a_m $ $ \forall m$ and using the general form of an incident field given by \eqref{-49a}, the near-field source amplitudes can be obtained as
\beq{5}
\left.
 \ba
\Phi_d &=\sum\limits_{n=-\infty}^\infty E_{n}^{(p)} U_n^{\,+}(k_p{\bf x}),
\\&\\
\Psi_d &= \sum\limits_{n=-\infty}^\infty E_{n}^{(s)} U_n^{\,+}(k_s{\bf x}),
\ea
\right\}
\quad
\ \ \text{ for }
\ \  |{\bf x}|< \text{max}(|{\bf x}_m|-a_m),
\eeq
where
 \bal{6}
E_{n}^{(p)} =\sum\limits_{m=1}^M  \sum\limits_{l=-\infty}^\infty B_{m,\,l}^{(p)} V_{n-l}^{\,-}({k_p\bf x}_m),
\quad
E_{n}^{(s)} =\sum\limits_{m=1}^M  \sum\limits_{l=-\infty}^\infty B_{m,\,l}^{(s)} V_{n-l}^{\,-}(k_s{\bf x}_m). 
 \eal
If the total field is zero in the near-field, then we must have $E_{n}^{(p)}+ A_n^{(p)}$ and $E_{n}^{(s)}+A_n^{(s)}$ ensuring the necessity of \eqref{7}$_{3,4}$. Sufficiency is guaranteed by substituting the expressions \eqref{7}$_{3,4}$ into an assumed near-field of the form \eqref{5}.

\subsection{Divergence of the active field summation} \label{sec:div}

The infinite sum expression for the active source fields defined by \eqref{-49b} with source amplitudes \eqref{-9a}-\eqref{-9c} is formally valid only in $|{\bf x}-{\bf x}_m|>a_m$. That is, the expression is not itself valid in the domain $\mathcal{A}$ in which the sources {reside}! A valid form could be obtained by using the alternative version of Graf's addition theorem in the domain $\mathcal{A}_m$ associated with the arc $\pa C_m$ but the usual form of Graf in the domain $\mathcal{A}_m$ associated with all other $\pa{C}_n, n\neq m$. We would then be assured that the active field is zero everywhere outside $\C$. However if we were to do this, the $m$th source would not be present in the domain $\mathcal{A}_m$ since the active field would be bounded by construction.

Active cloaking therefore requires that we use the expression \eqref{-49b} with source amplitudes \eqref{-9a}-\eqref{-9c} for the active field everywhere but we must take a finite number of terms in the multipole expansion. That is,  we use the source amplitudes that appear in the infinite sum as motivation for the choice of source amplitudes that should be chosen in an active field that contains only a finite number of multipoles. This ensures a finite (but large) field inside $\mathcal{A}$. We should note that this type of difficulty and the fact that it may be used to our advantage in the anti-sound context was noted by Kempton \cite{Kem-76}.

With a finite sum for the active field therefore, the integral equation \eqref{5=6} is not perfectly satisfied but instead
\begin{align}
\mathbf{u}({\bf x}) \approx \begin{cases}
0, & {\bf x}\in \C, \\
\mathbf{u}_i({\bf x}), & {\bf x}\notin{\C\cup\mathcal{A}},
\end{cases}
\end{align}
and the field is large (but finite) inside $\mathcal{A}$. Finally we note that a straightforward truncation of the active field may not be optimal in terms of cloaking and ensuring a non-radiating field. This issue will be considered elsewhere.

\section{Numerical examples} \label{sec4}

\begin{figure}[htb]	
	\centering
		\includegraphics[width=3in]{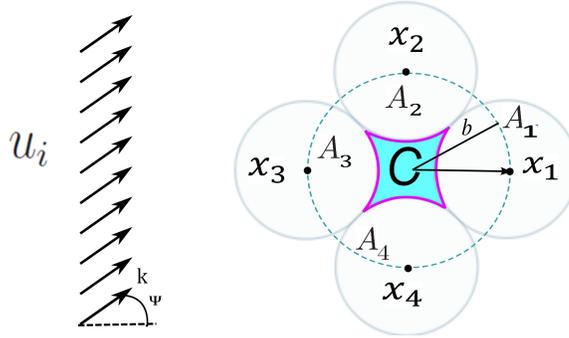}  
\caption{Plane wave insonification of the cloaking region $\C$ generated by four ($M = 4$) active sources placed at the corners of a square.}
\label{fig:fig_2}
\end{figure}

The numerical calculations for  active source configurations of the type shown in Fig.\ \ref{fig:fig_2} are performed for plane longitudinal and transverse incident waves of a unit amplitude,  $(A_p = 1,\ A_s = 0)$ and $(A_p = 0,\ A_s = 1)$,
  for angles of incidence $\psi_p=\psi_s=7^o $.  Variable values are taken for the wavenumbers $k_p$ and $k_s$, the number of sources $M$, and the number of terms $N$ in summations \eqref{-4a} and \eqref{6} (the truncation size).  The  $M$ active sources are  symmetrically located on a circle of radius $b$,  with
\beq{-236}
a_m =a, \quad
|{\bf x}_m|=b,\quad
\beta_m 
= (m-1)\beta_0, \quad  \beta_0= 2\pi\Big( \frac{m-1}M  \Big),
\quad m=\overline{1,  M},
\eeq
where  $\beta_m$ is the argument of vector ${\bf x}_m$, and $a\ge   b \sin\frac{\pi} M$. The  circular arcs are defined by
\beq{122}
\theta^{(m)}_{1,2} = \pi + \beta_m \mp
\bigg|
\sin^{-1} \left( \frac ba \sin \frac{\pi} M
\right) - \frac{\pi} M  \bigg|,
\quad m=\overline{1,  M}.
\eeq
In all examples, we take $ a =   b \sin\frac{\pi} M$ and consider an elastic medium having a property of aluminum with $c_p= 6427\,m/s,\,c_s=3112\,m/s,\,\rho= 2694\, kg/m^3$  \cite{Honarvar97}.

\subsection{The scattering amplitudes}

\begin{figure}[htb]	
	\centering
		\includegraphics[width=4.5in]{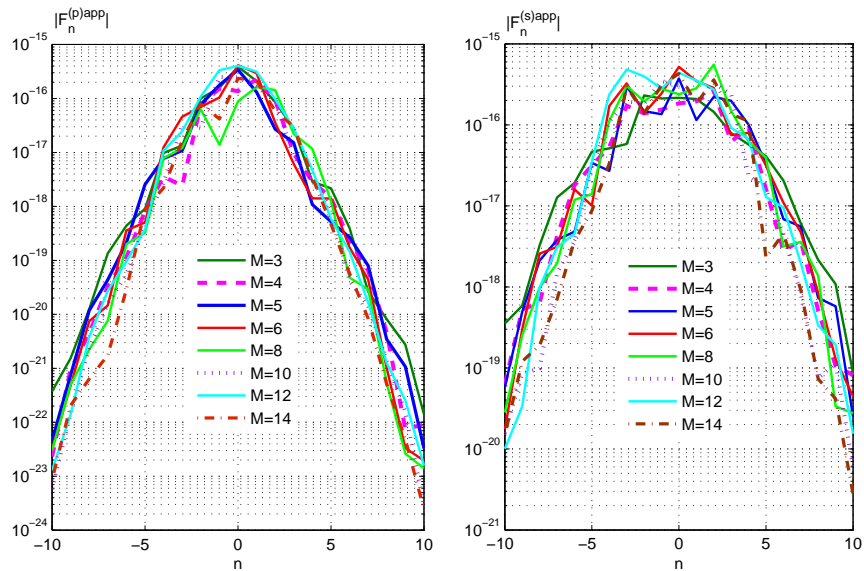}    
\caption{Variation of the far-field amplitude coefficients with number of active sources
 $(M=\overline{3,14})$ for transverse incident waves.  In all cases $N=100$, $\psi_s=7^o $. }
\label{fig:Farfield_S}
\end{figure}

Consider the truncated versions of the infinite sums in eq. \eqref{6} for the farfield amplitudes $F_n^{(p)}$ and $F_n^{(s)}$, and eq. \eqref{-4a}  for the nearfield  amplitudes $E_n^{(p)}$ and $E_n^{(s)}$:
\begin{subequations}\label{14X}
\bal{14Xa}
\left.
\begin{matrix}
F_n^{(p)app}
 \\
E_n^{(p)app}
\end{matrix}
\right\}
&=\sum\limits_{m=1}^M\sum\limits_{l=-N}^N
B_{m,l}^{(p)} \,	
\times
\begin{cases}
V_{n-l}^{\,-}({\bf x}_m),   
\\
U_{n-l}^{\,-}({\bf x}_m)  ,
 \end{cases}
 	\quad \forall \, n\in \mathbb{Z} ,
 	\\
 \left.
\begin{matrix}
F_n^{(s)app}
 \\
E_n^{(s)app}
\end{matrix}
\right\}
&=\sum\limits_{m=1}^M\sum\limits_{l=-N}^N
B_{m,l}^{(s)} \,	
\times
\begin{cases}
V_{n-l}^{\,-}({\bf x}_m),   
\\
U_{n-l}^{\,-}({\bf x}_m)  ,
 \end{cases}
 	\quad \forall \, n\in \mathbb{Z} . \label{14Xb}
\eal
\end{subequations}

\begin{figure}[htb]	
	\centering
		\includegraphics[width=4.5in]{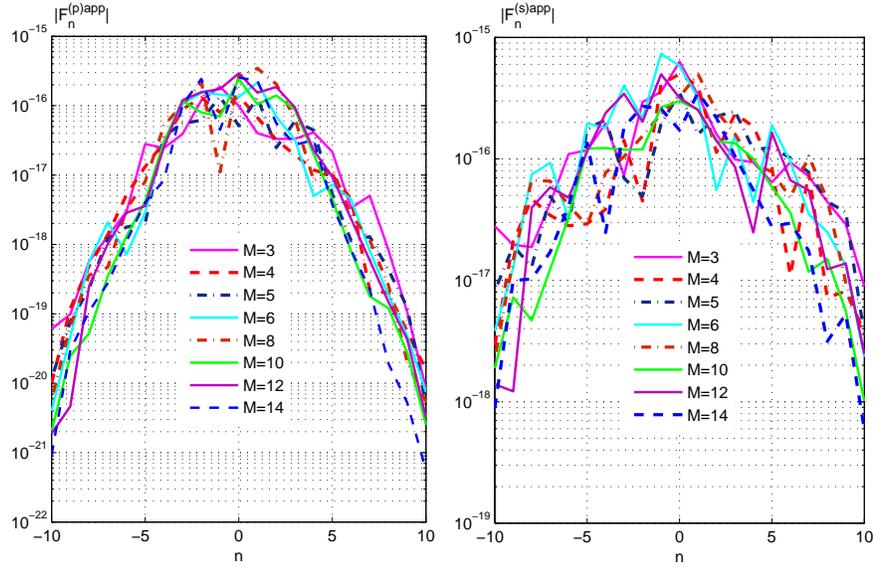}    
\caption{Variation of the far-field amplitude coefficients with number of active sources
 $(M=\overline{3,14})$ for longitudinal incident waves.  In all cases $N=100$, $\psi_s=7^o $. }
\label{Farfield_P}
\end{figure}

The approximate near-field $E_n^{(p)app}$, $E_n^{(s)app}$  and far-field $F_n^{(p)app}$, $F_n^{(s)app}$ amplitudes are calculated at the incident shear wavenumber $k_s=5$ varying the number of active sources $M$ and the truncation size $N$. The dependence of the far-field coefficients $|F_n^{(p)app}|$, $|F_n^{(s)app}|$
is illustrated in Fig.\ \ref{fig:Farfield_S} for transverse and in Fig.\ \ref{Farfield_P} for longitudinal  wave incidences.  As M increases, the far-field coefficients fluctuate at small $|n|$, and decrease at larger values of $|n|$ for both compressional and shear incident waves.

\begin{figure}[htb]	
	\centering
		\includegraphics[width=4.5in]{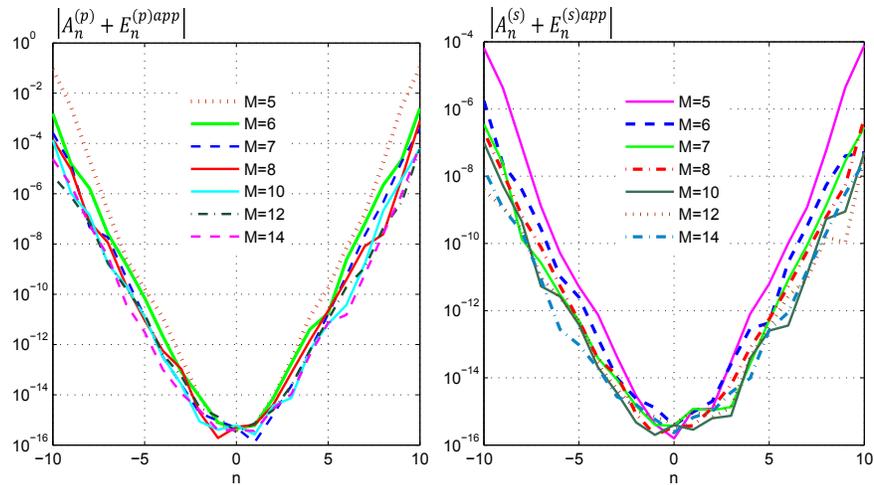}    
\caption{Dependence of the near-field amplitude coefficients on $n$, the order of Bessel function, varying the number of active sources
 $(M=\overline{5,14})$ for {transverse} wave incidence.  In all cases $N=100$, $\psi_s=7^o $. }
\label{fig:Nearfield_S}
\end{figure}

\begin{figure}[htb]	
	\centering
		\includegraphics[width=4.5in]{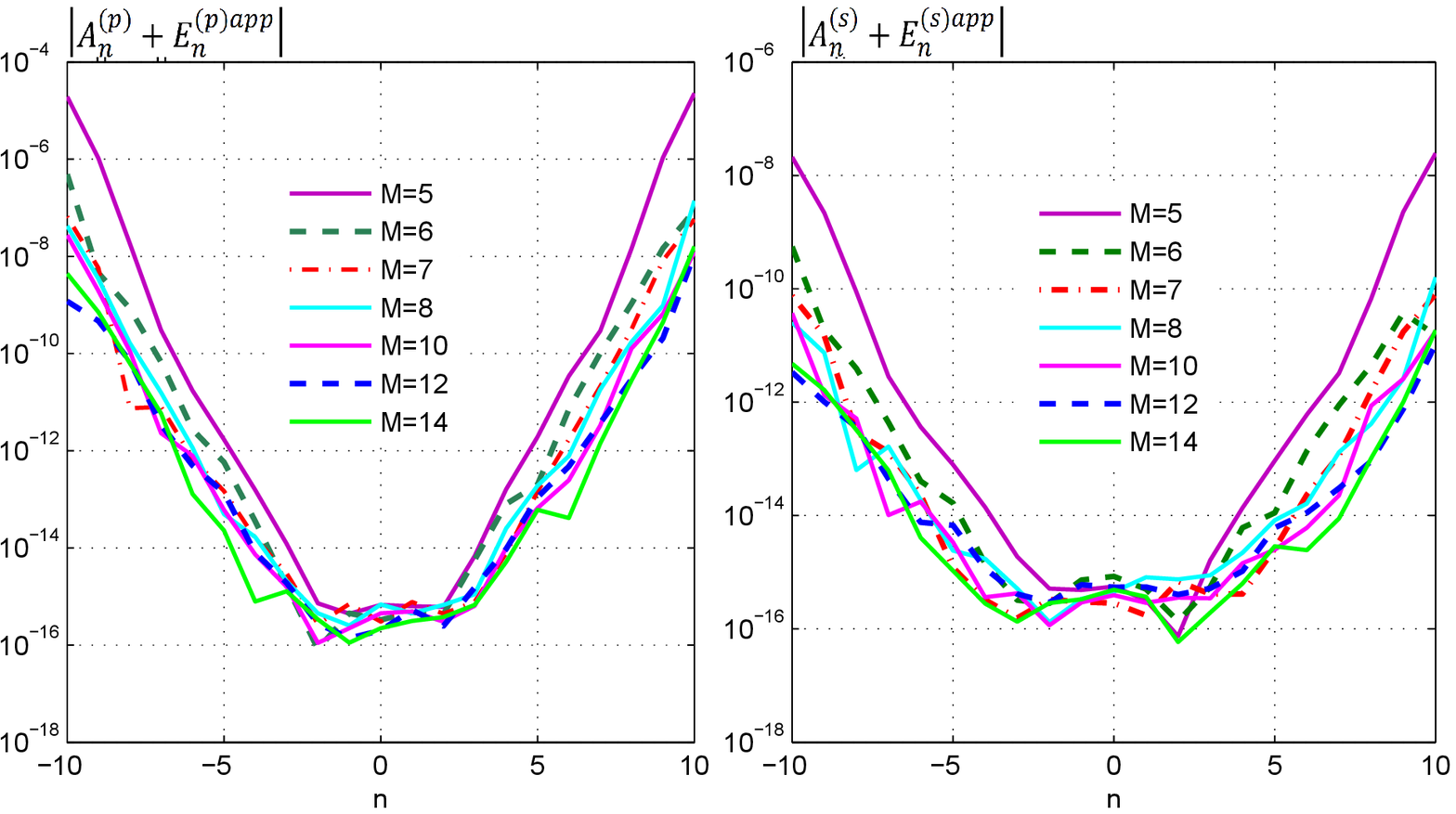}    
\caption{Variation of the near-field amplitude coefficients with number of active sources
 $(M=\overline{5,14})$ for longitudinal incident waves.  In all cases $N=100$, $\psi_s=7^o $. }
\label{Nearfield_P}
\end{figure}

The variation of the near-field coefficients  $|A_n^{(p)} + E_n^{(p)app}|$, $|A_n^{(s)} + E_n^{(s)app}|$  with the number of sources $M$ is depicted in Fig.\ \ref{fig:Nearfield_S} for transverse and in Fig.\ \ref{Nearfield_P} for longitudinal incident waves. For longitudinal wave incidence, the near-field  $|A_n^{(p)}+E_n^{(p)app}|$ is less than $10^{-4}$ and $|A_n^{(s)} + E_n^{(s)app}|$ is less than $10^{-7}$. On the contrary, the  results are less accurate for transverse waves, as the near-field $|A_n^{(p)}+E_n^{(p)app}|$  approaches the order of $10^{-1}$ and $|A_n^{(s)}+E_n^{(s)app}|$ reaches a value $10^{-4}$.

\begin{figure}[htb]	
	\centering
		\includegraphics[width=4.5in]{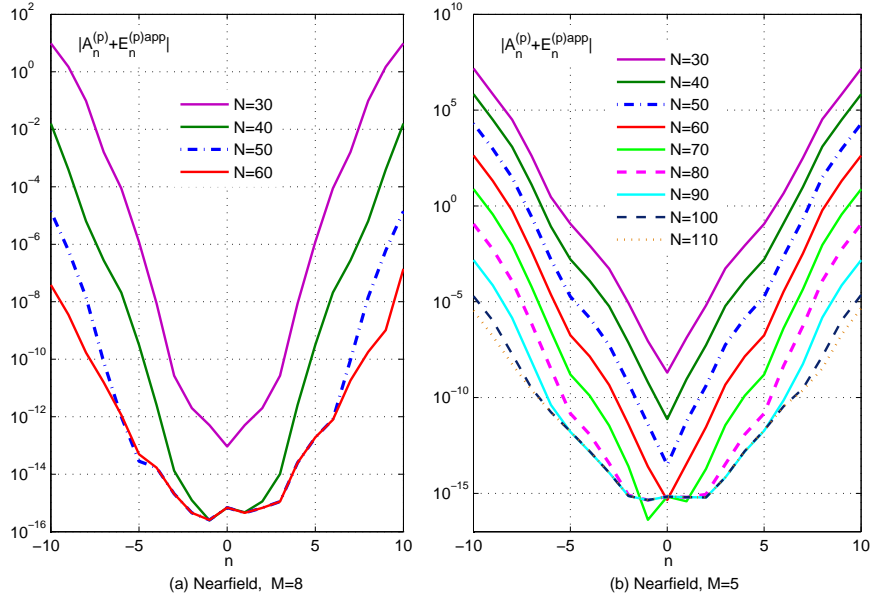}    
\caption{The near-field amplitude coefficients as a function of $n$, the order of Bessel function for different values of the truncation size $N$ in \eqref{14X} generated by different numbers of active sources: (a) $M=5$   and (b) $M=8$, for longitudinal incident waves.  }
\label{fig:Nearfield_P_N}
\end{figure}

Figure \ref{fig:Nearfield_P_N} displays the near-field amplitude coefficients $|A_n^{(p)}+E_n^{(p)app}|$  and $|A_n^{(s)} + E_n^{(s)app}|$ as functions of $n$, the order of Bessel function, for different values of  $N$ and $M$. The accuracy of the near-field coefficients improves as $N$ and $M$ increase. Increasing the number of sources $M$ allows a  decrease in the truncation size $N$ and the order of error.

\subsection{Total displacement field}\label{sec4_2}

\subsubsection{Longitudinal plane wave incidence}\label{subsec4_2_1}

\begin{figure}[ht] \centering
 \subfigure[$N=5$]{
  \includegraphics[scale=0.5]{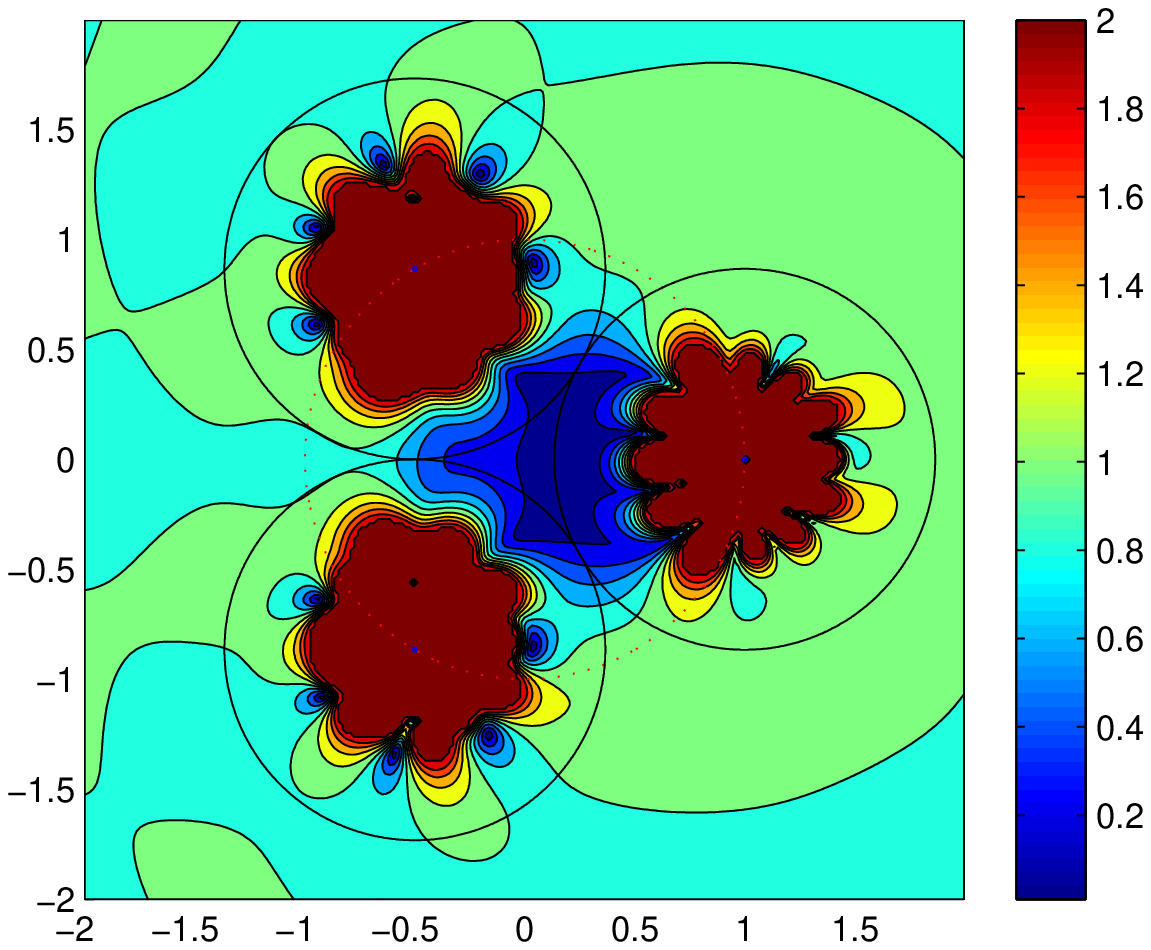}
   \label{fig:subfig201}
   }
 \subfigure[$N=10$]{
  \includegraphics[scale=0.5]{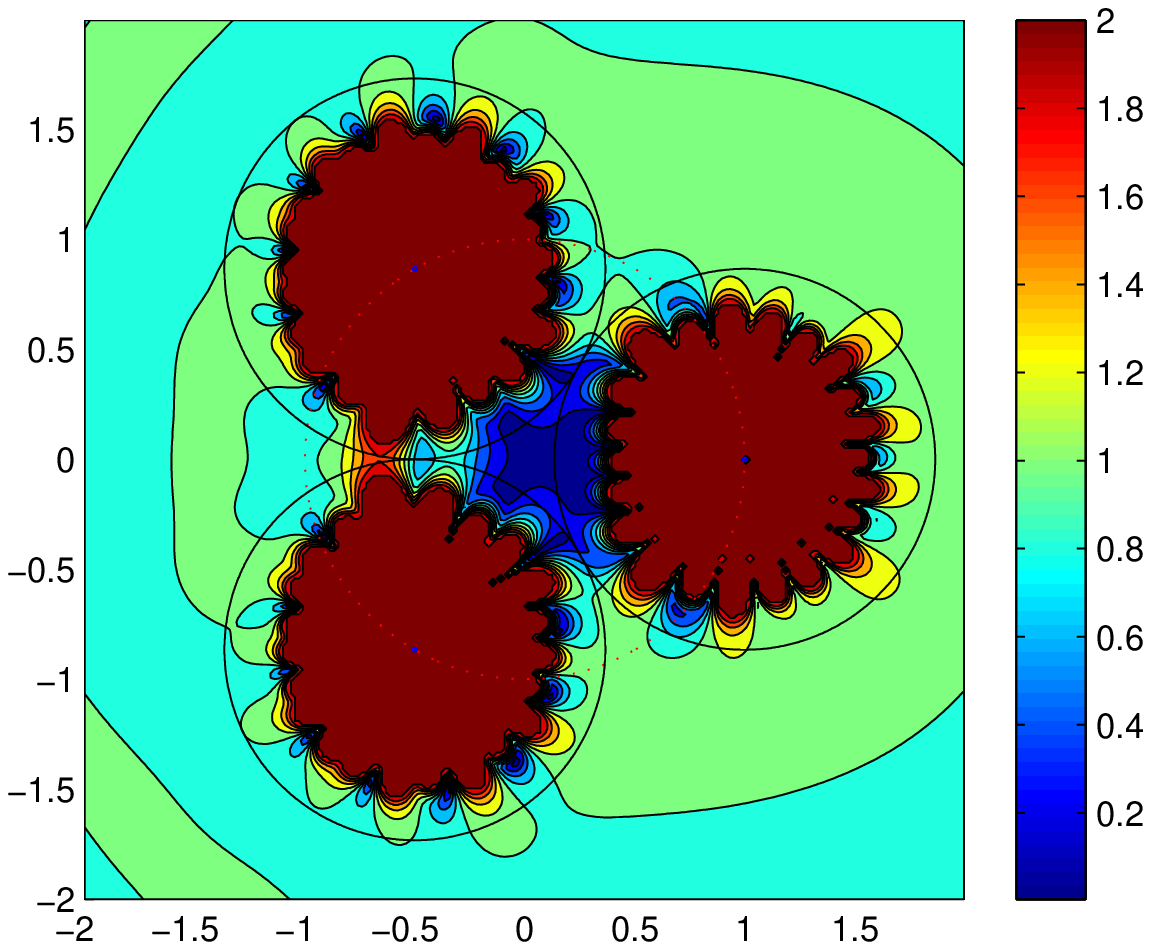}
   \label{fig:subfig202}
   }
 \subfigure[$N=20$]{
  \includegraphics[scale=0.5]{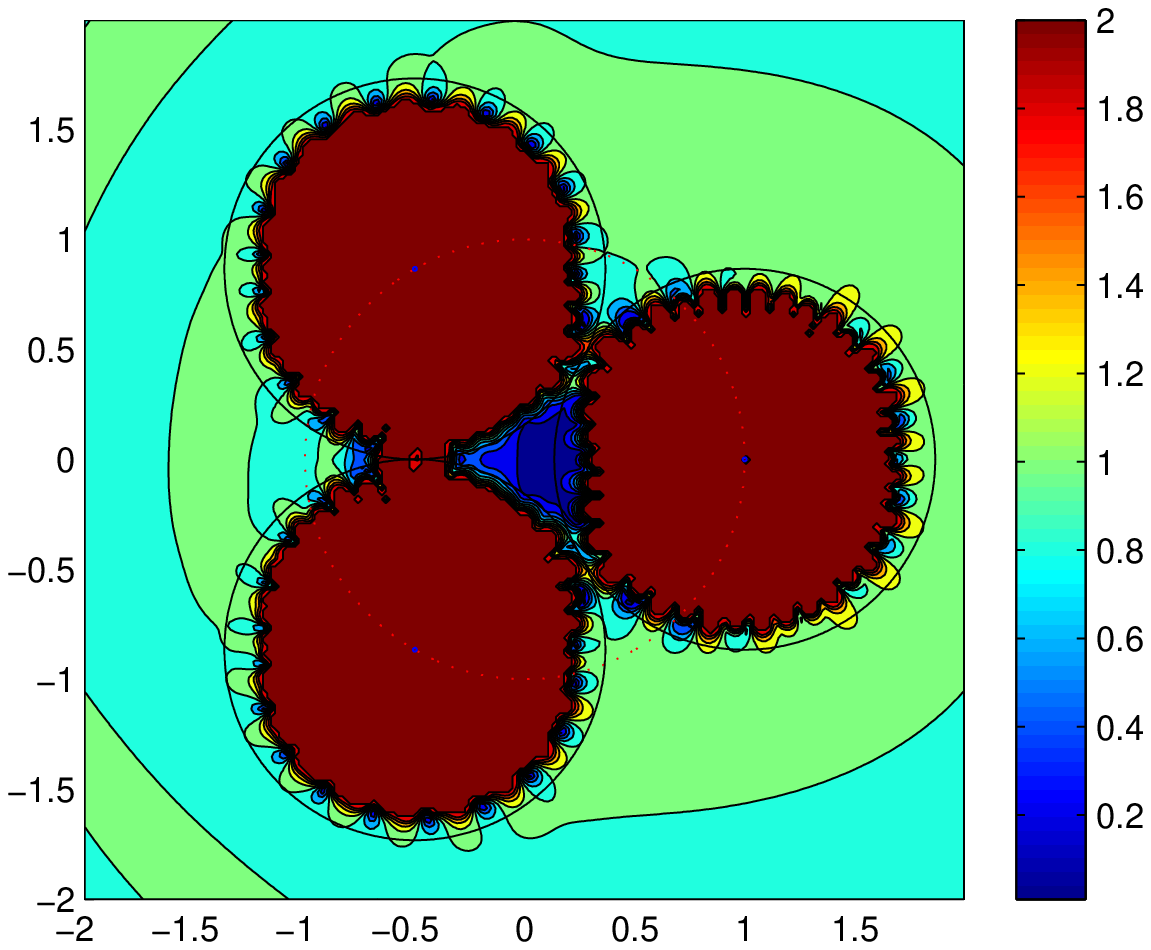}
   \label{fig:subfig203}
   }
 \subfigure[$N=50$]{
  \includegraphics[scale=0.5]{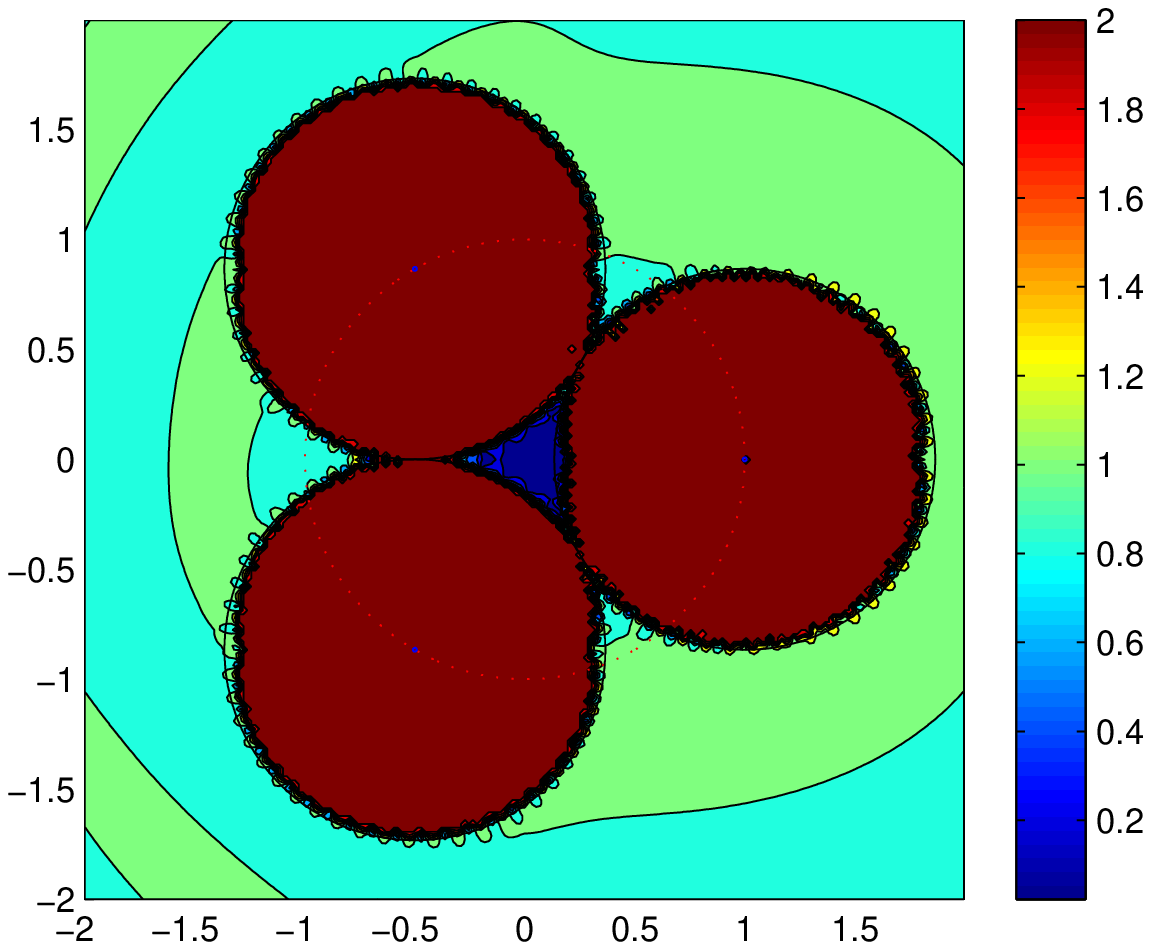}
   \label{fig:subfig204}
   }
 \caption[]{%
  Absolute  value of displacement vector component  $|u_x|/k_p$ for $N=5$ \subref{fig:subfig201}, $N=10$ \subref{fig:subfig202}, $N=20$ \subref{fig:subfig203}  and  $N=50$ \subref{fig:subfig204}  when cloaking devices are  active with $M=3, k_p =2$ for longitudinal wave incidence.}
 \label{fig:fig_20}
\end{figure}

\begin{figure}[ht]
 \centering
 \subfigure[$N=5$]{
  \includegraphics[scale=0.5]{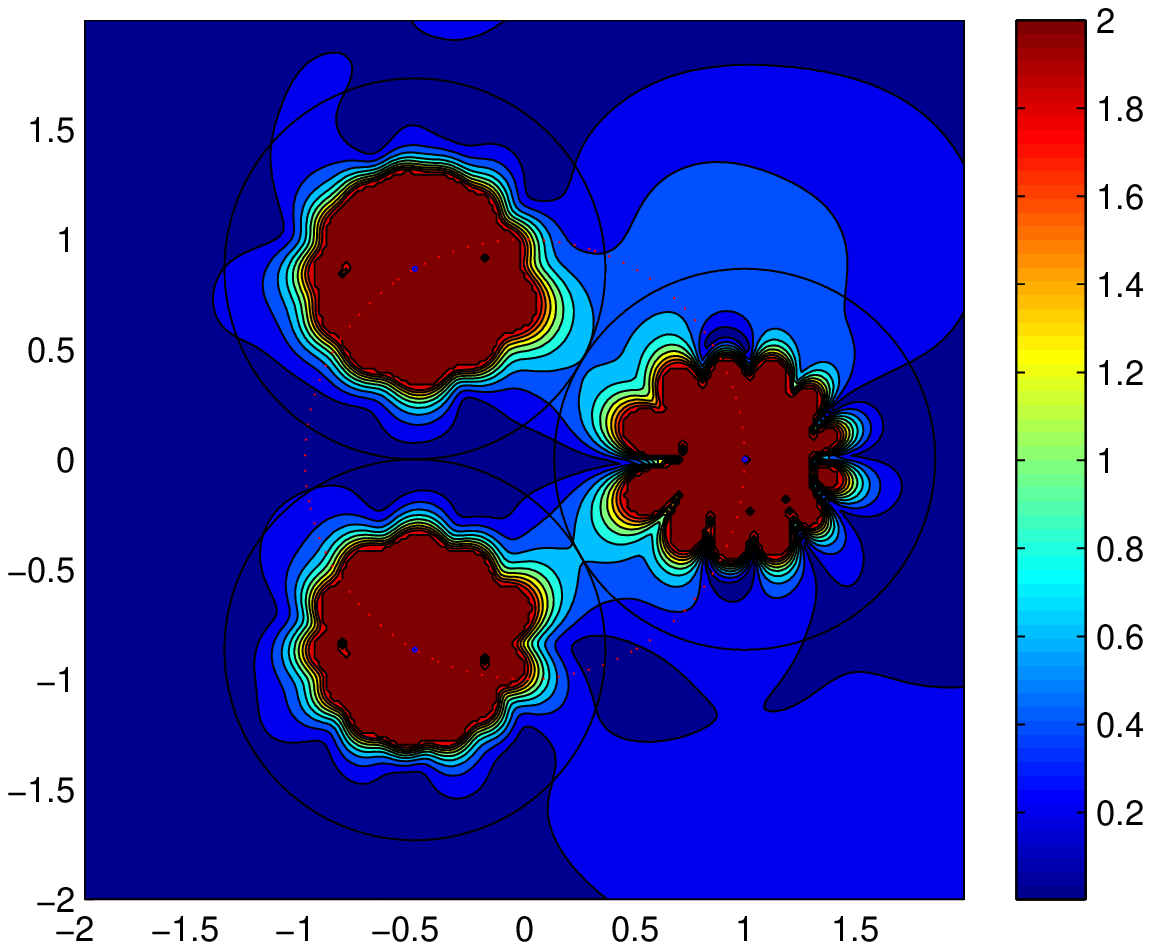}
   \label{fig:subfig31}
   }
 \subfigure[$N=10$]{
  \includegraphics[scale=0.5]{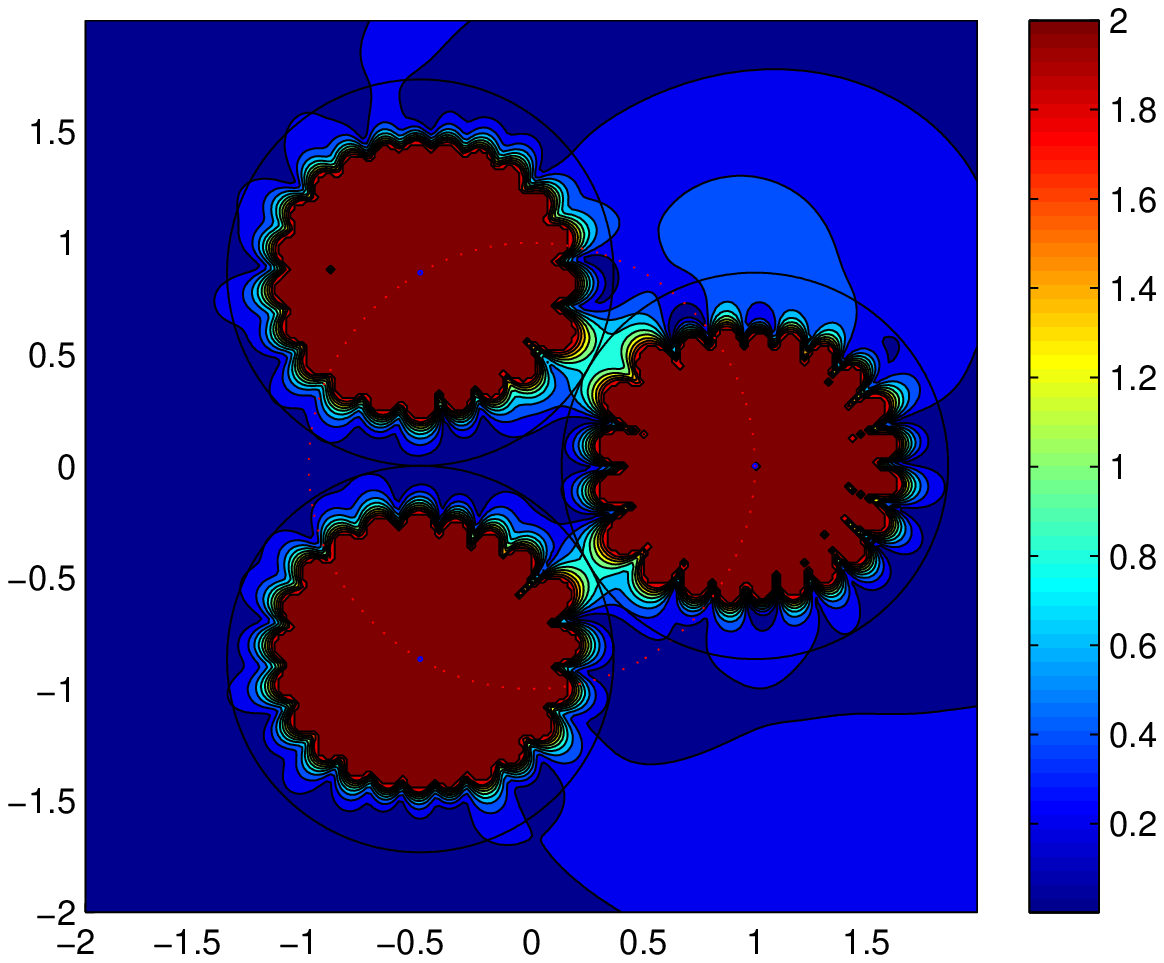}
   \label{fig:subfig32}
   }
 \subfigure[$N=20$]{
  \includegraphics[scale=0.5]{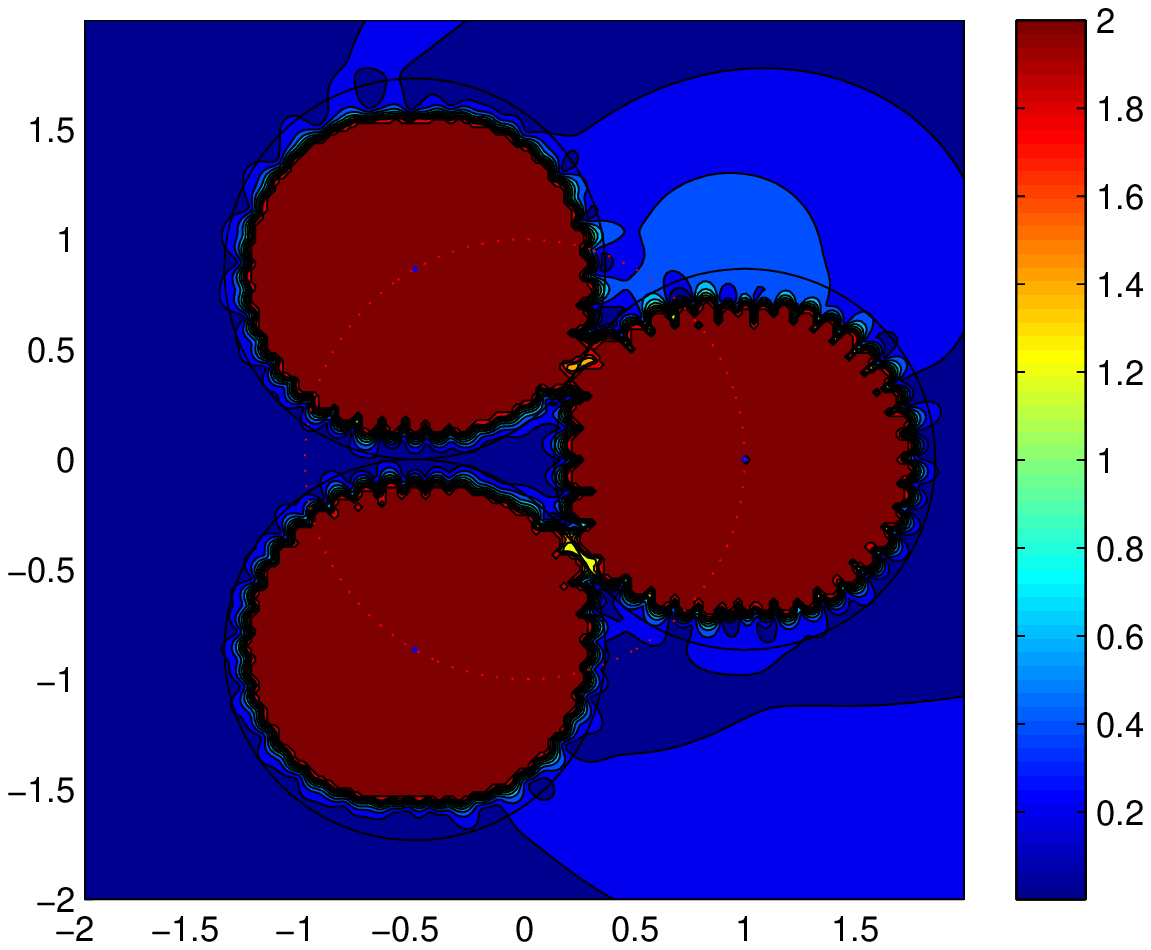}
   \label{fig:subfig33}
   }
 \subfigure[$N=30$]{
  \includegraphics[scale=0.5]{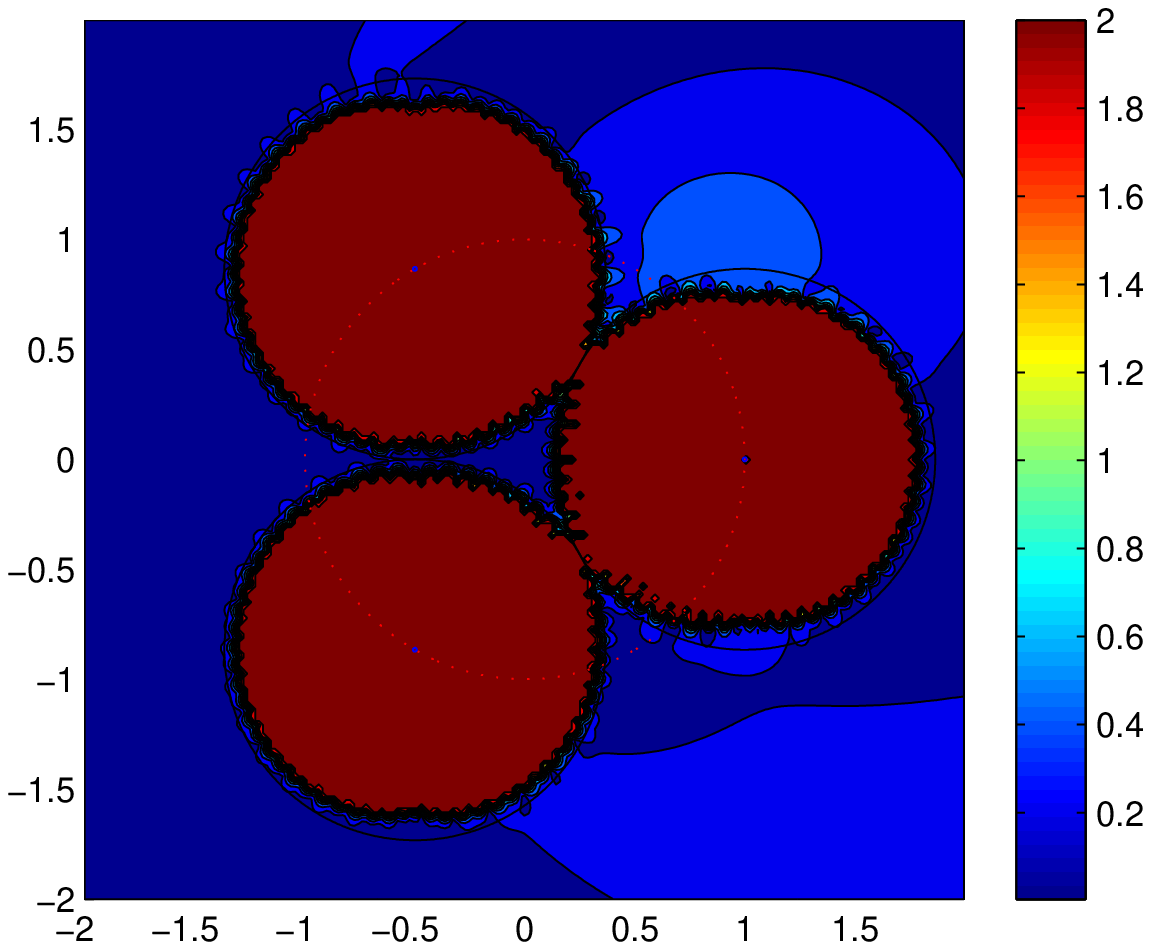}
   \label{fig:subfig34}
   }
  \caption[]{%
  Absolute  value of displacement vector components $|u_y|/k_p$  for $N=5$ \subref{fig:subfig31}, $N=10$ \subref{fig:subfig32}, $N=20$ \subref{fig:subfig33}  and  $N=30$ \subref{fig:subfig34}  when cloaking devices are  active with $M=3, k_p =2$ for longitudinal wave incidence.}
\label{fig:fig_30}
\end{figure}

First, consider longitudinal plane wave incidence of the form \eqref{4=23}. The total displacement vector components in Cartesian coordinates are
\beq{3d}
\big( u_x,\, u_y\big)  =   \Big( \frac{\partial { \Phi_i} }{\partial x}+\frac{\partial { \Phi_d} }{\partial x}+\frac{\partial { \Psi_d} }{\partial y}, \, \frac{\partial { \Phi_i} }{\partial y}+\frac{\partial { \Phi_d} }{\partial y}-\frac{\partial { \Psi_d} }{\partial x} \Big).
\eeq

Introducing eq. \eqref{4=23} and \eqref{-49b} into \eqref{3d} yields
\begin{subequations}\label{5d}
\bal{5da}
\frac{u_x}{k_p} &=    \sum\limits_{m=1}^M\sum\limits_{n=-\infty}^\infty
  \Bigg[   B_{m,n}^{(p)} \bigg(   \cos\theta_m   {V_n^+}^\prime  \big(k_p({\bf x}-{\bf x}_m) \big)
  -i n \sin\theta_m   \frac{ {V_n^+}  \big(k_p({\bf x}-{\bf x}_m) \big)}{k_p|{\bf x}-{\bf x}_m|} \bigg)
   \notag
\\ &
+ B_{m,n}^{(s)} \bigg(  \kappa \sin \theta_m   {V_n^+}^\prime  \big(k_s({\bf x}-{\bf x}_m) \big)
+ i n \cos \theta_m   \frac{ {V_n^+}  \big(k_s({\bf x}-{\bf x}_m) \big)}{k_p|{\bf x}-{\bf x}_m|} \bigg)
\Bigg]+ i \cos \psi_p \Phi_i,
\\
\frac{u_y}{k_p} &=   \sum\limits_{m=1}^M\sum\limits_{n=-\infty}^\infty \Bigg[
B_{m,n}^{(p)}  \bigg(  \sin\theta_m   {V_n^+}^\prime  \big(k_p({\bf x}-{\bf x}_m) \big)
+i n \cos\theta_m   \frac{ {V_n^+}  \big(k_p({\bf x}-{\bf x}_m) \big) }{k_p|{\bf x}-{\bf x}_m|} \bigg)
\notag
\\&
 + B_{m,n}^{(s)} \bigg( - \kappa \cos \theta_m   {V_n^+}^\prime  \big(k_s({\bf x}-{\bf x}_m)\big)
 + i n \sin\theta_m   \frac{ {V_n^+}  \big(k_s({\bf x}-{\bf x}_m) \big)}{k_p|{\bf x}-{\bf x}_m|} \bigg)  \Bigg] + i \sin \psi_p \,  \Phi_i ,
 \label{5db}
\eal
\end{subequations}
where
\beq{6d}
\theta_m  ({\bf x}) = \arg ({\bf x}-{\bf x}_m).
\eeq

\subsubsection{Transverse plane wave incidence}\label{subsec4_2_2}

Transverse incident plane waves are of the form \eqref{4=30}. The total displacement vector components in Cartesian coordinates are
\beq{11d}
\big( u_x,\, u_y\big)  =   \Big( \frac{\partial { \Phi_d} }{\partial x}+\frac{\partial { \Psi_i} }{\partial y}+\frac{\partial { \Psi_d} }{\partial y}, \, \frac{\partial { \Phi_d} }{\partial y}-\frac{\partial { \Psi_i} }{\partial x}-\frac{\partial { \Psi_d} }{\partial x} \Big).
\eeq

Introducing eq. \eqref{4=30} and \eqref{-49b} into \eqref{11d} yields
\begin{subequations}\label{13d}
\bal{13da}
\frac{u_x}{k_s} &=     \sum\limits_{m=1}^M\sum\limits_{n=-\infty}^\infty
  \Bigg[ B_{m,n}^{(p)} \bigg(   \kappa^{-1} \cos\theta_m   {V_n^+}^\prime  \big(k_p({\bf x}-{\bf x}_m) \big)
  - i  n \sin\theta_m   \frac{ {V_n^+}  \big(k_p({\bf x}-{\bf x}_m) \big)}{k_s|{\bf x}-{\bf x}_m|} \bigg) \notag
\\
&
+ B_{m,n}^{(s)} \bigg(   \sin \theta_m   {V_n^+}^\prime  \big(k_s({\bf x}-{\bf x}_m) \big)
+i n \cos \theta_m   \frac{ {V_n^+}  \big(k_s({\bf x}-{\bf x}_m) \big)}{k_s|{\bf x}-{\bf x}_m|}
 \bigg) \Bigg] + i \sin \psi_s \Psi_i,
\\
\frac{u_y}{k_s} &= \sum\limits_{m=1}^M\sum\limits_{n=-\infty}^\infty \Bigg[ B_{m,n}^{(p)}  \bigg(  \kappa^{-1} \sin\theta_m   {V_n^+}^\prime  \big(k_p({\bf x}-{\bf x}_m) \big)     + i n \cos\theta_m   \frac{ {V_n^+}  \big(k_p({\bf x}-{\bf x}_m) \big) }{k_s|{\bf x}-{\bf x}_m|} \bigg)   \notag
\\
&
 + B_{m,n}^{(s)} \bigg( - \cos \theta_m   {V_n^+}^\prime  \big(k_s({\bf x}-{\bf x}_m)\big) + i n \sin\theta_m   \frac{ {V_n^+}  \big(k_s({\bf x}-{\bf x}_m) \big)}{k_s|{\bf x}-{\bf x}_m|} \bigg)  \Bigg] - i \cos \psi_s \,  \Psi_i ,
 \label{13db}
\eal
\end{subequations}
where $\theta_m$ is defined by \eqref{6d}.

\subsubsection{Results} \label{subsec4_2_3}

\begin{figure}[ht]
 \centering
 \subfigure[$k_p=5, M=3$]{
  \includegraphics[scale=0.5]{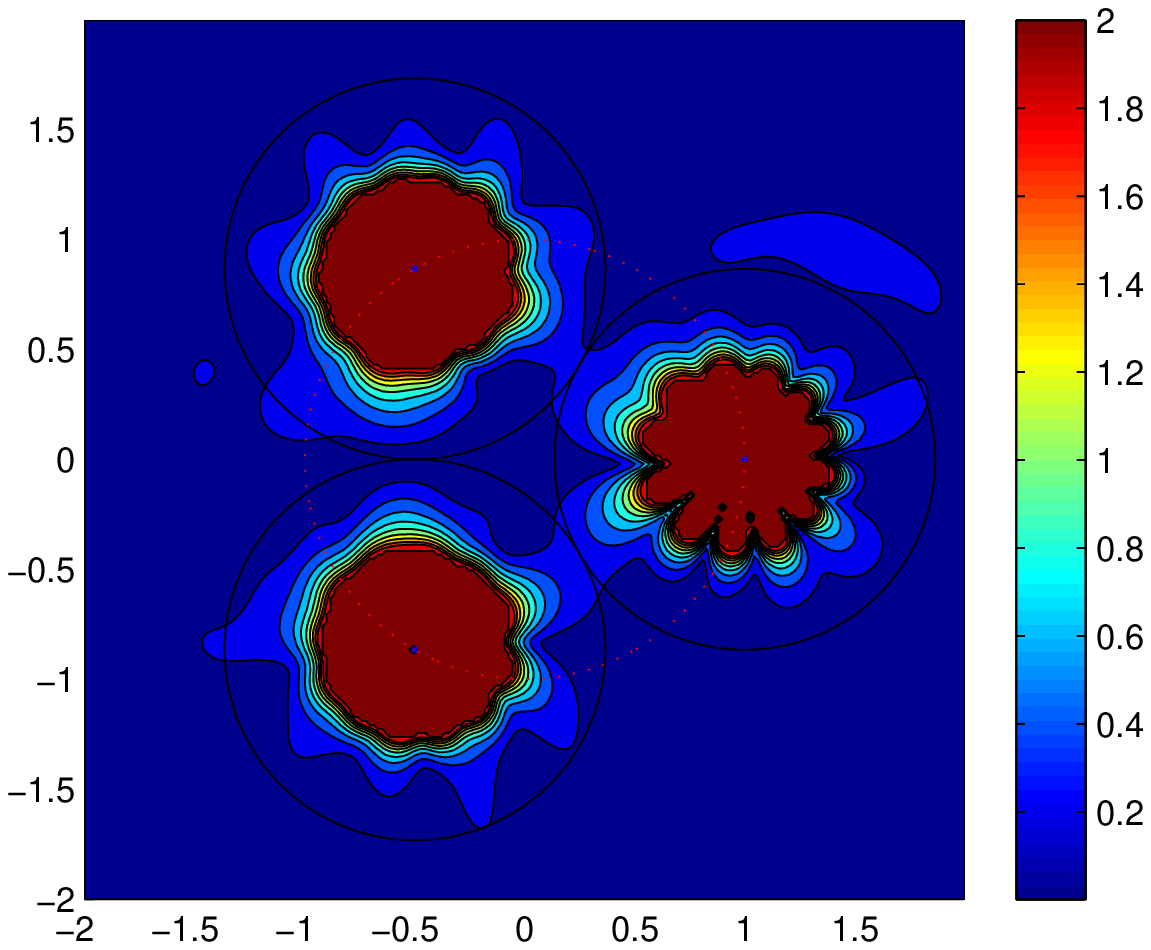}
   \label{fig:subfig41}
   }
 \subfigure[$k_p=10, M=3$]{
  \includegraphics[scale=0.5]{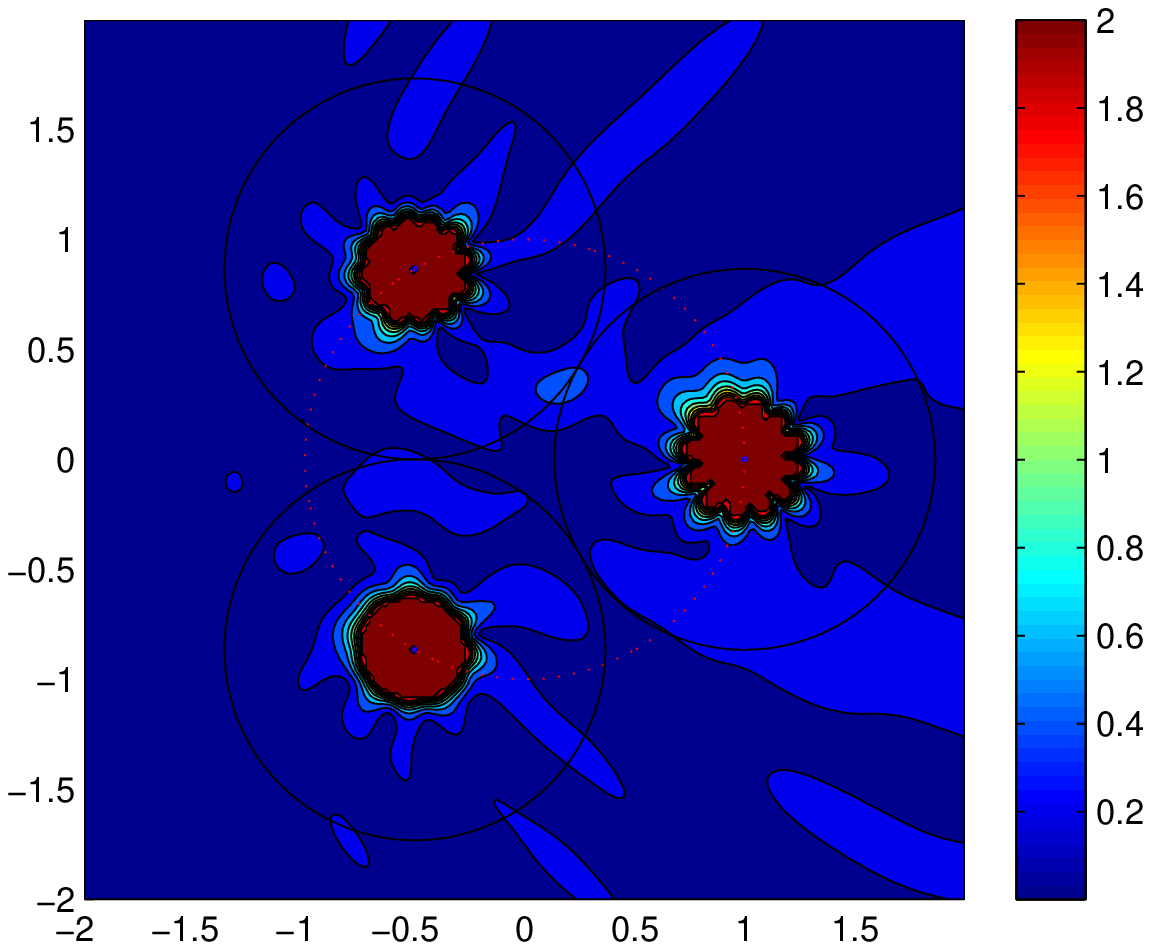}
   \label{fig:subfig42}
   }
 \subfigure[$k_p=5, M=7$]{
  \includegraphics[scale=0.5]{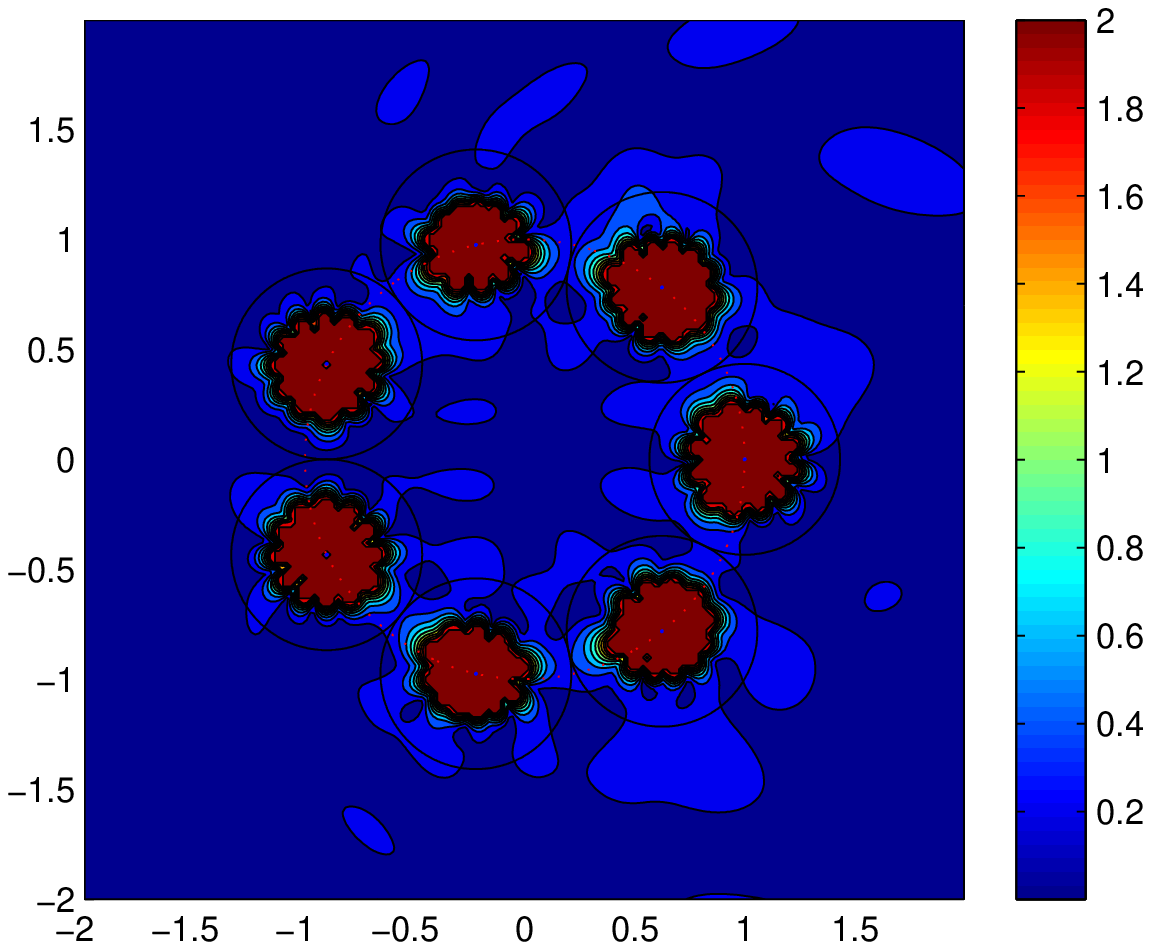}
   \label{fig:subfig43}
   }
 \subfigure[$k_p=10, M=7$]{
  \includegraphics[scale=0.5]{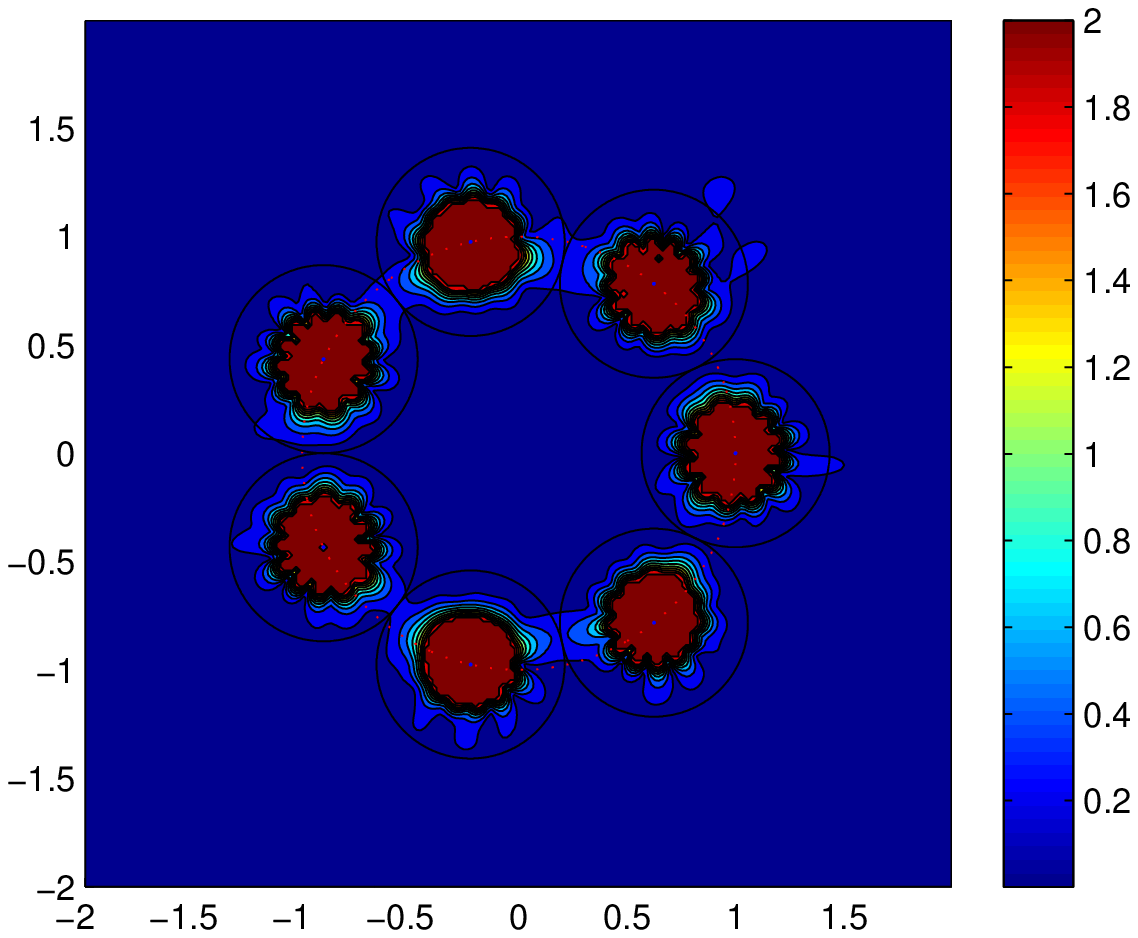}
   \label{fig:subfig44}
   }
  \caption[]{%
  Absolute  value of displacement vector component  $|u_y|/k_p$  for $k_p=10, M=3$ \subref{fig:subfig41}, $k_p=10, M=3$ \subref{fig:subfig42}, $k_p=5, M=7$ \subref{fig:subfig43}  and  $k_p=10, M=7$ \subref{fig:subfig44}  when cloaking devices are  active with $N=5$ for longitudinal wave incidence.}
\label{fig:fig_40}
\end{figure}

\begin{figure}[ht]
 \centering
 \subfigure[$N=5, M=3$]{
  \includegraphics[scale=0.5]{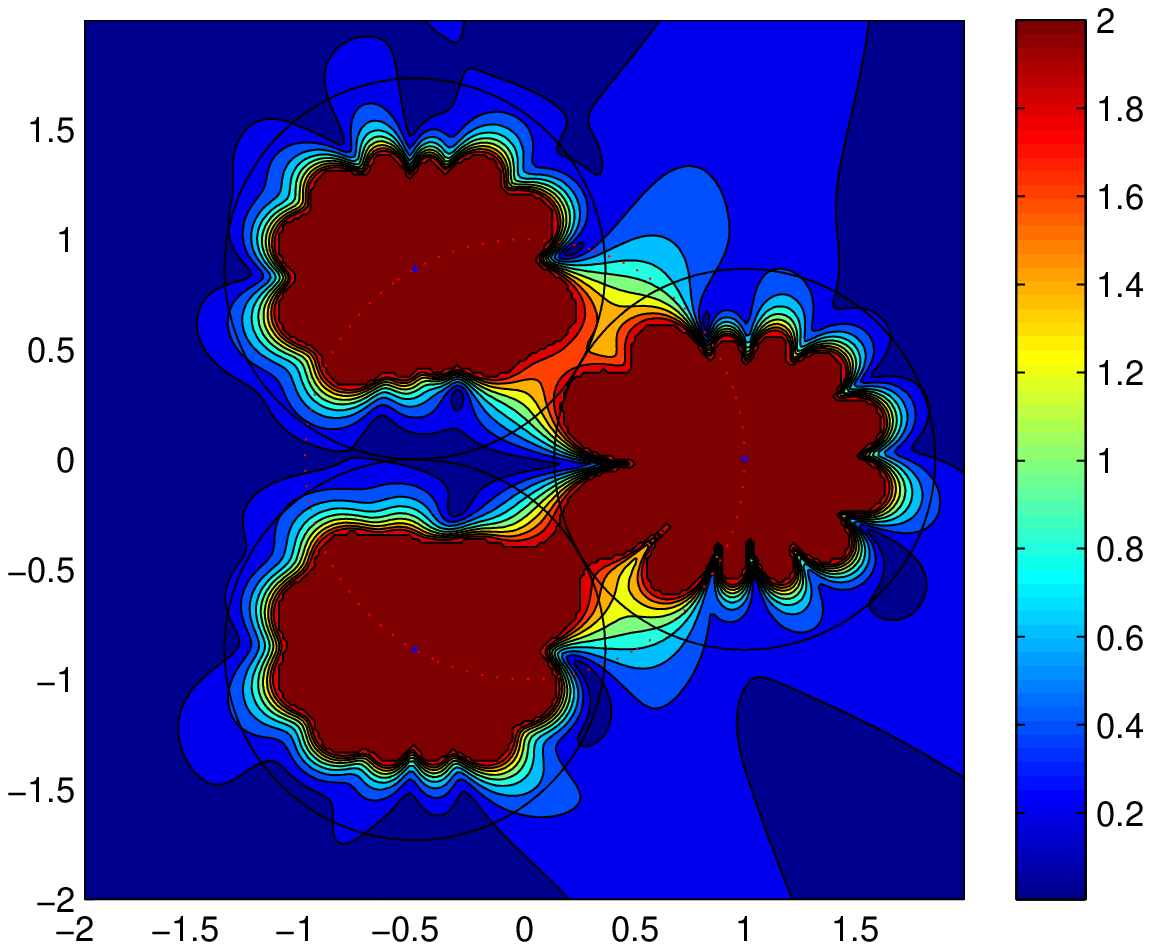}
   \label{fig:subfig51}
   }
 \subfigure[$N=50, M=3$]{
  \includegraphics[scale=0.5]{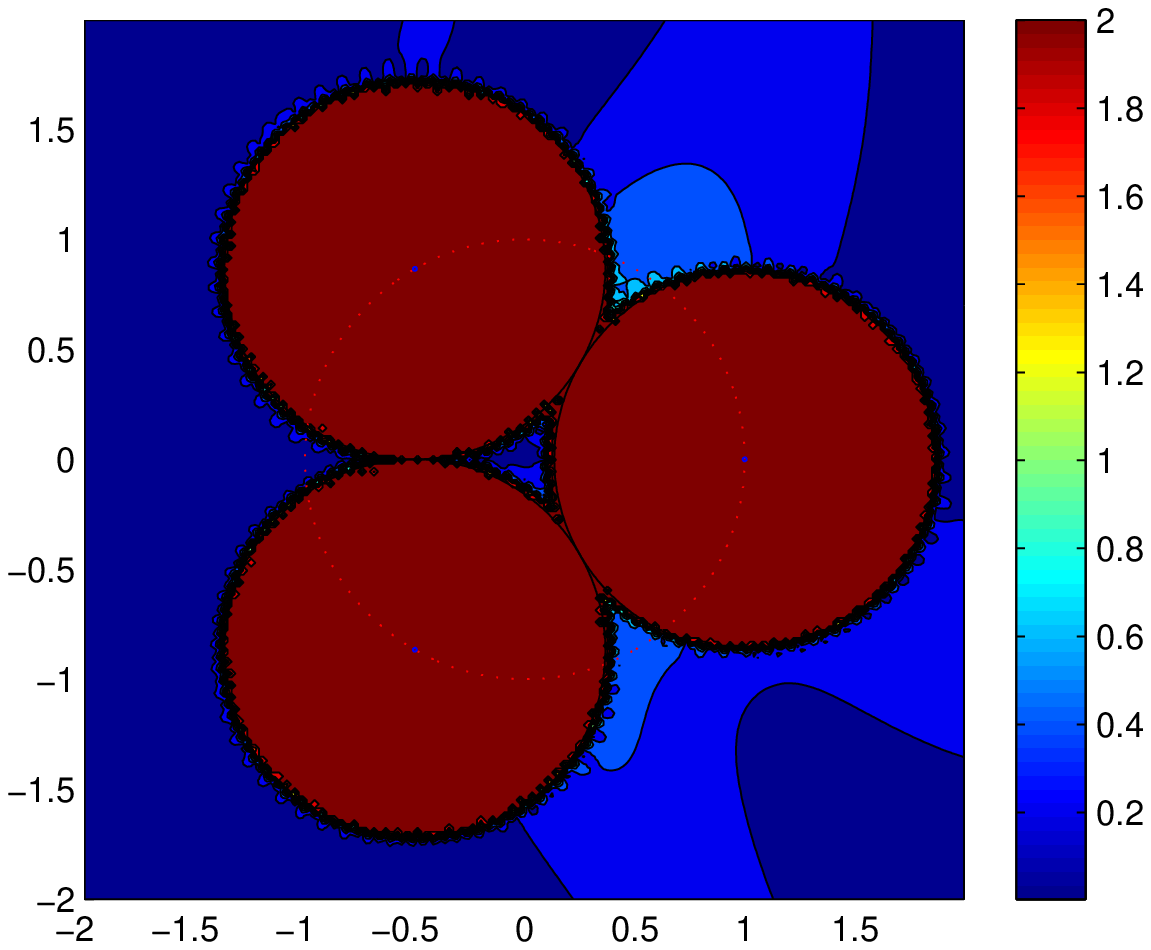}
   \label{fig:subfig52}
   }
 \subfigure[$N=5, M=6$]{
  \includegraphics[scale=0.5]{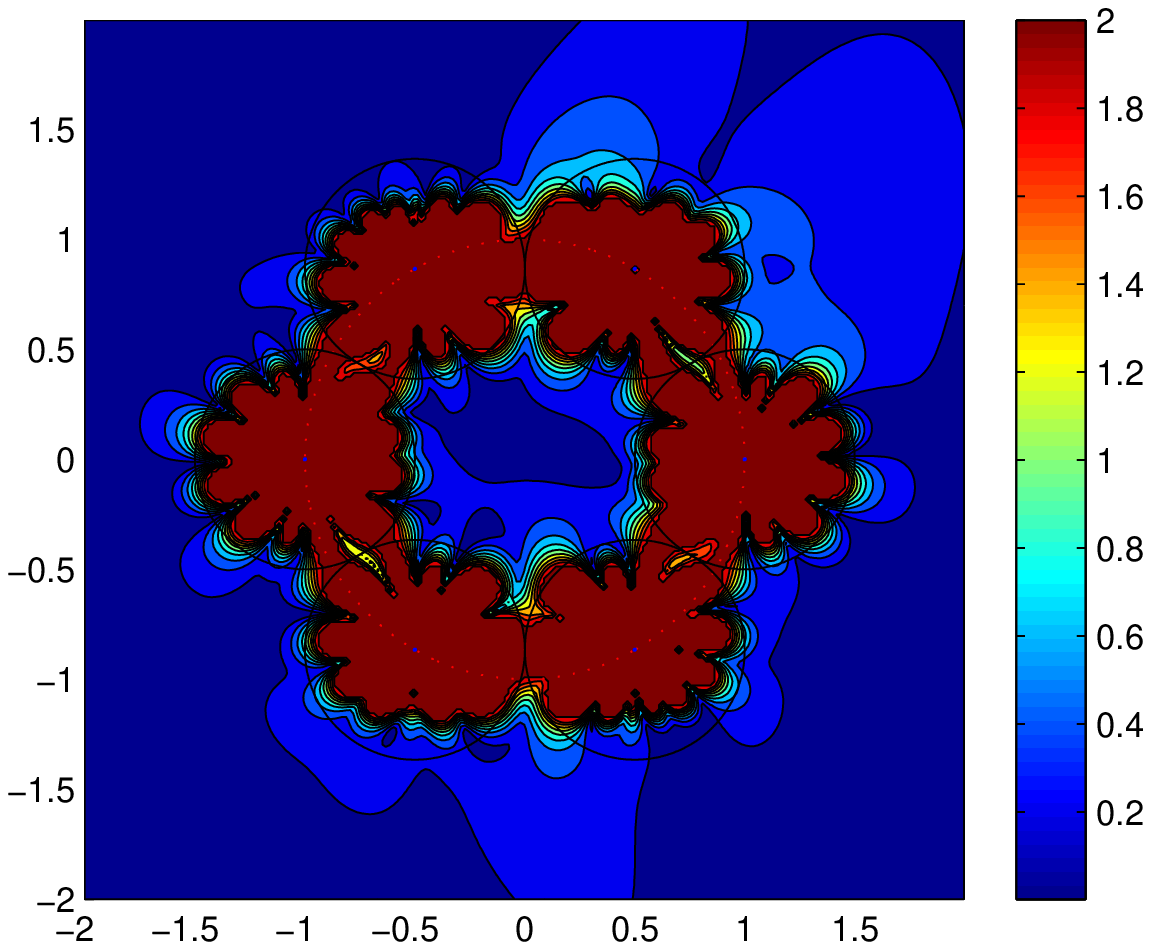}
   \label{fig:subfig53}
   }
 \subfigure[$N=50, M=6$]{
  \includegraphics[scale=0.5]{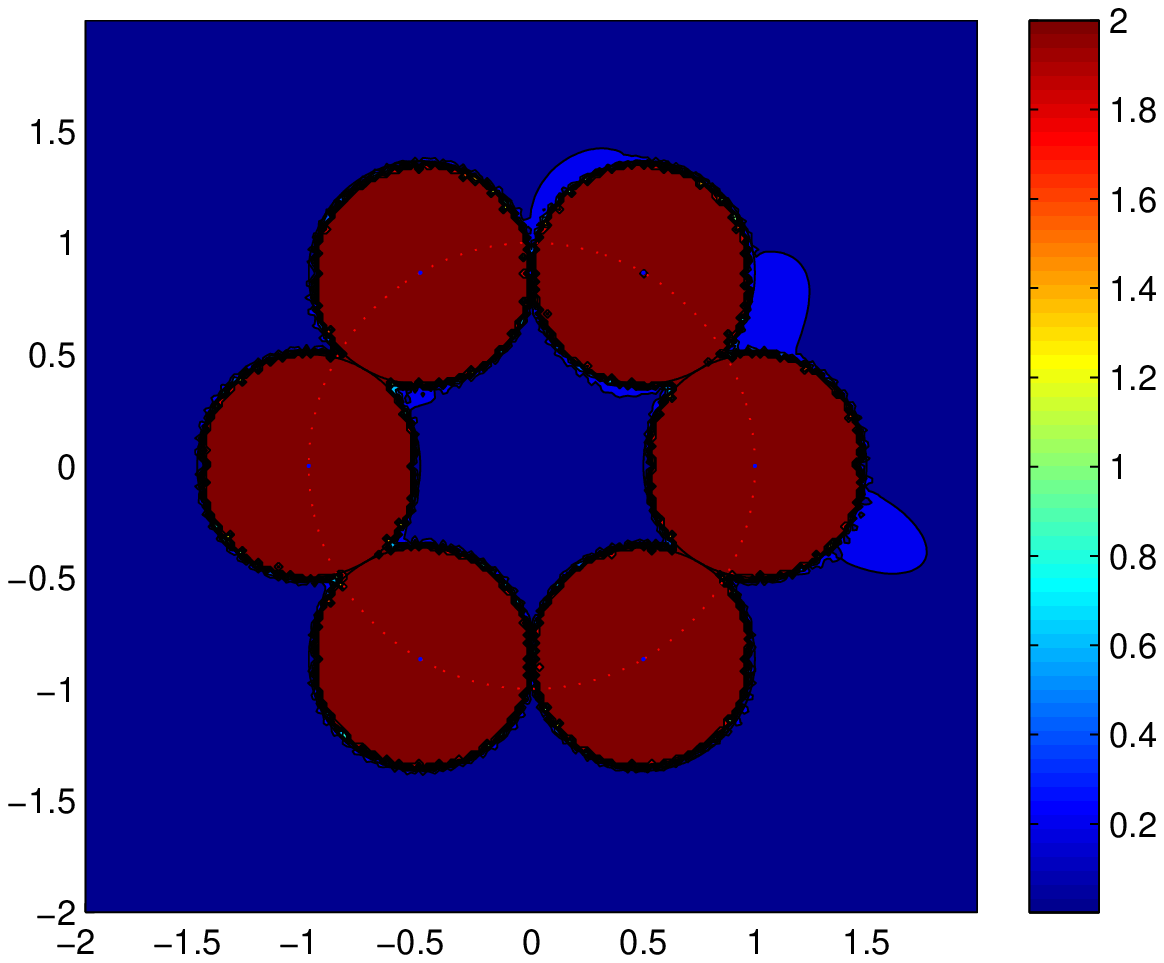}
   \label{fig:subfig54}
   }
  \caption[]{%
    Absolute  value of displacement vector component   $|u_x|/k_s$ for $N=5, M=3$ \subref{fig:subfig51}, $N=50, M=3$ \subref{fig:subfig52}, $N=5, M=6$ \subref{fig:subfig53}  and  $N=50, M=6$ \subref{fig:subfig54}  when cloaking devices are  active with $k_p =2, k_s= 4.1305$ for transverse wave incidence.}
\label{fig:fig_50}
\end{figure}

\begin{figure}[ht]
 \centering
 \subfigure[$N=5, k_p=2$]{
  \includegraphics[scale=0.5]{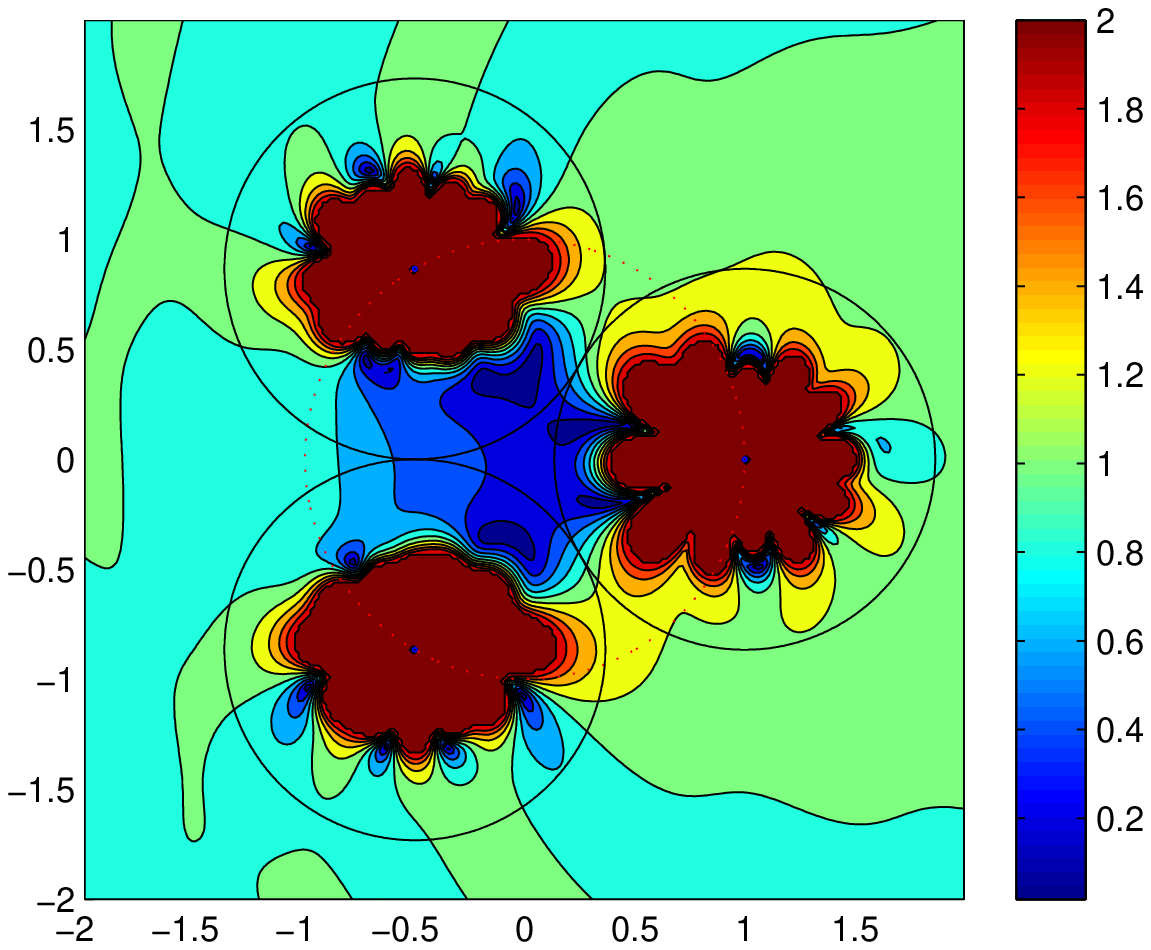}
   \label{fig:subfig61}
   }
 \subfigure[$N=5, k_p=5$]{
  \includegraphics[scale=0.5]{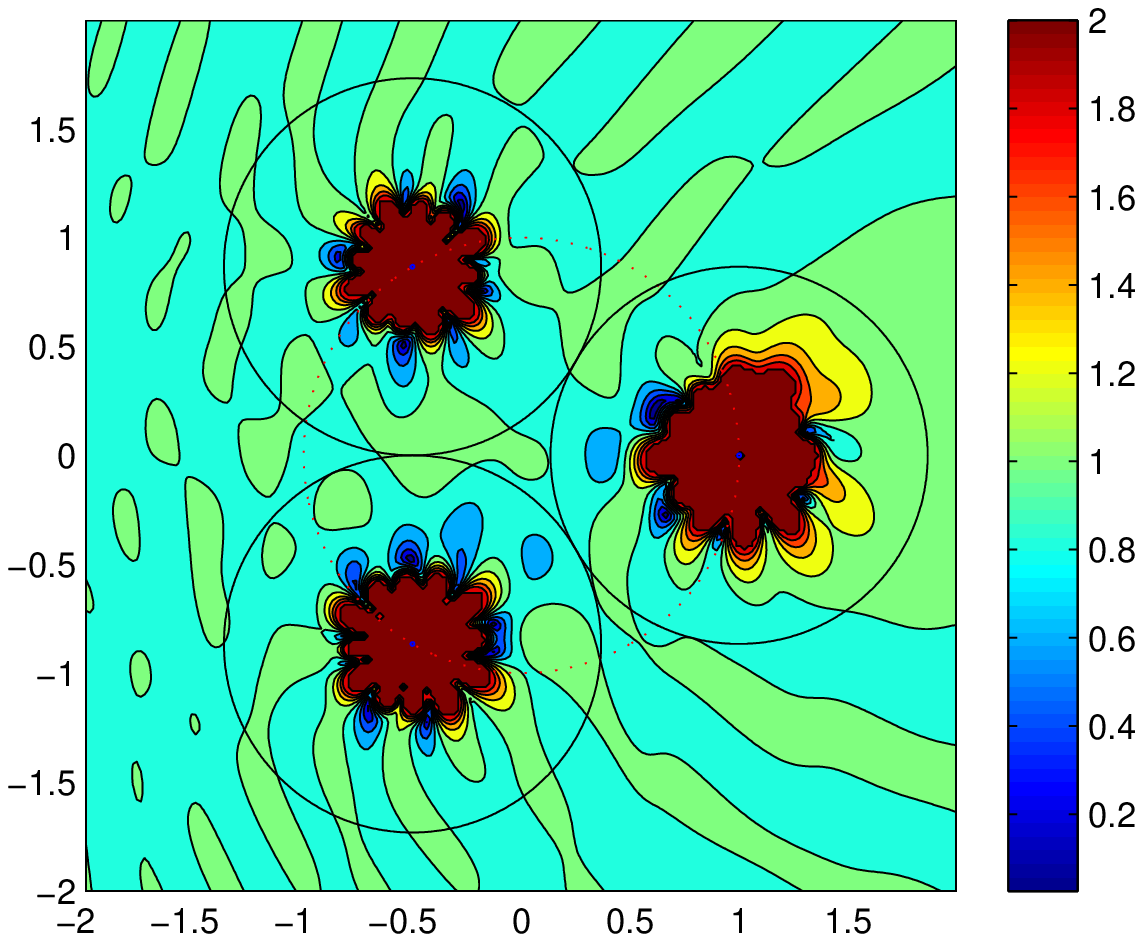}
   \label{fig:subfig62}
   }
 \subfigure[$N=20, k_p=2$]{
  \includegraphics[scale=0.5]{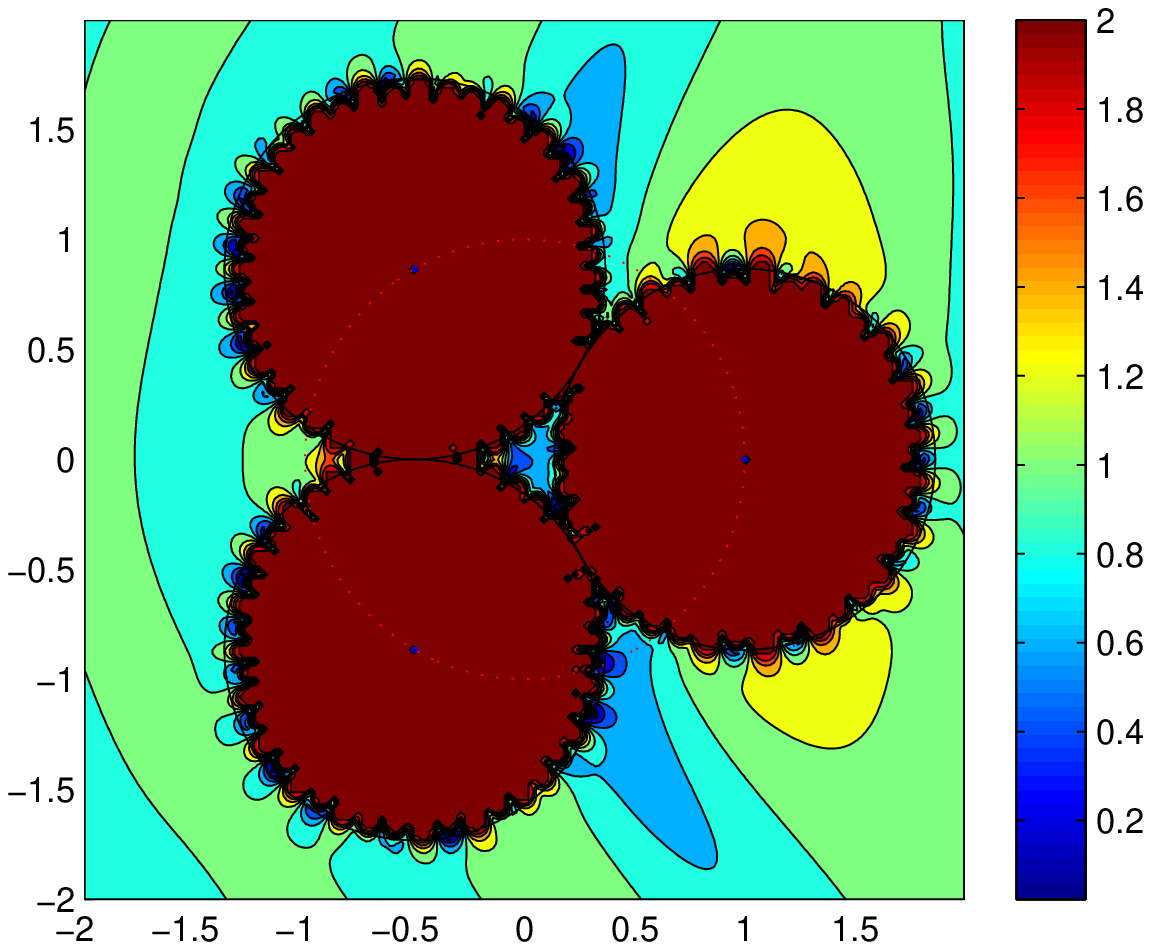}
   \label{fig:subfig63}
   }
 \subfigure[$N=20, k_p=5$]{
  \includegraphics[scale=0.5]{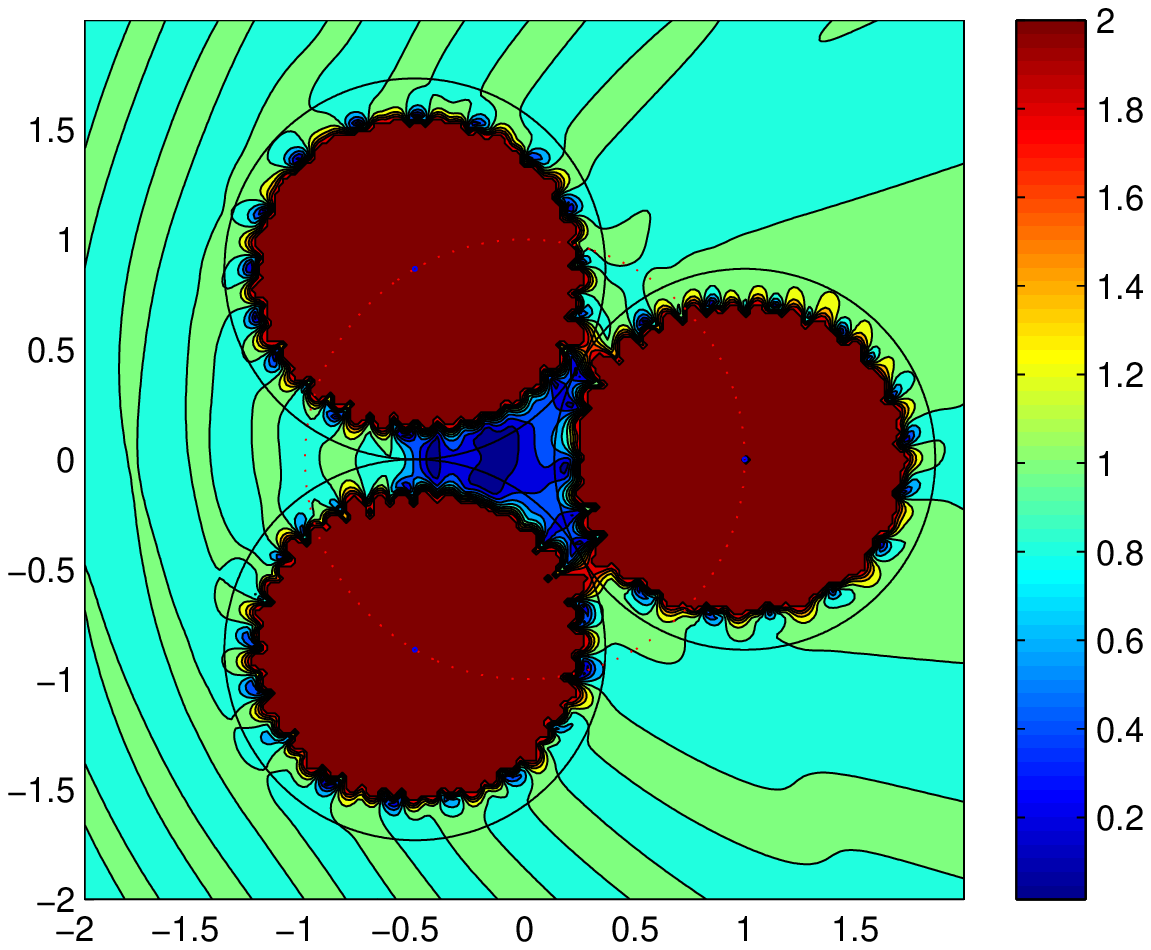}
   \label{fig:subfig64}
   }
   \subfigure[$N=50, k_p=2$]{
  \includegraphics[scale=0.5]{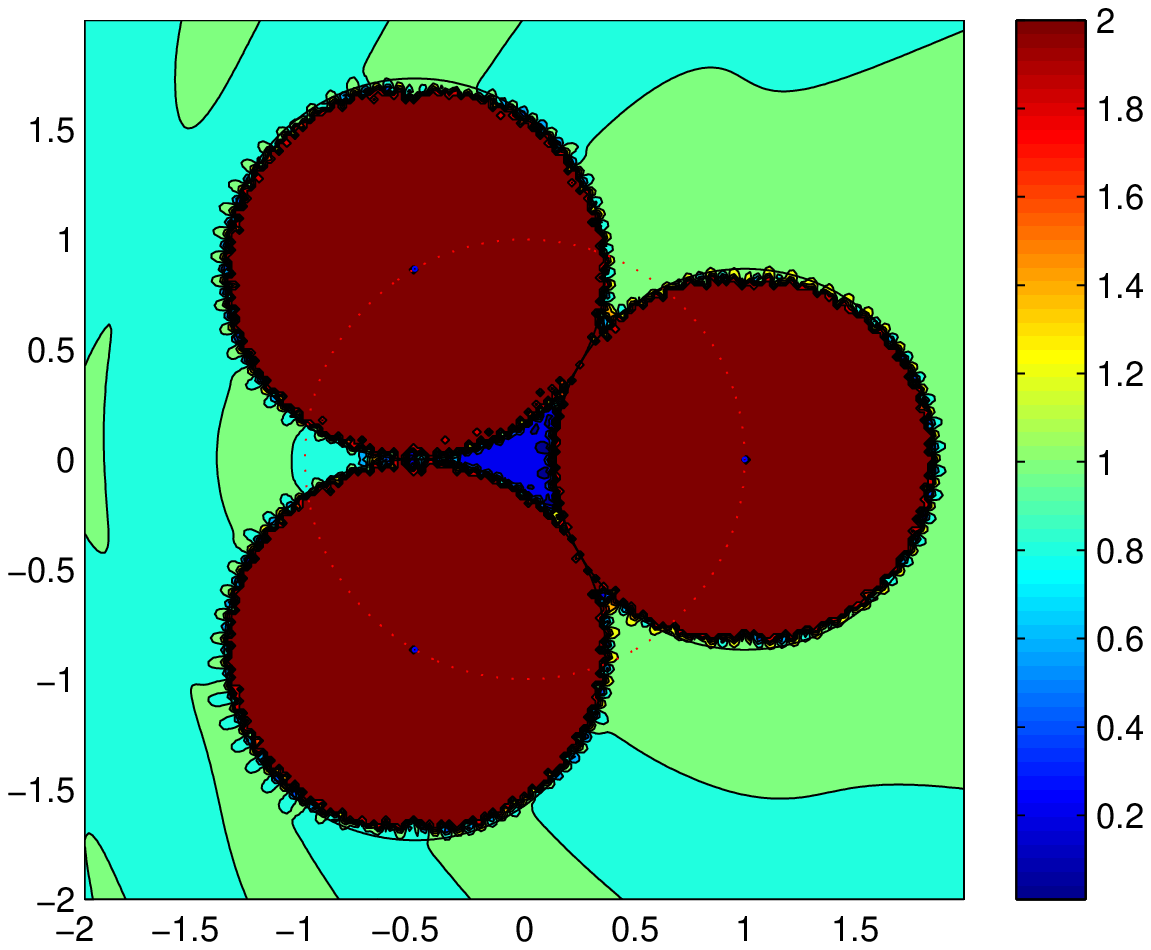}
   \label{fig:subfig65}
   }
 \subfigure[$N=50, k_p=5$]{
  \includegraphics[scale=0.5]{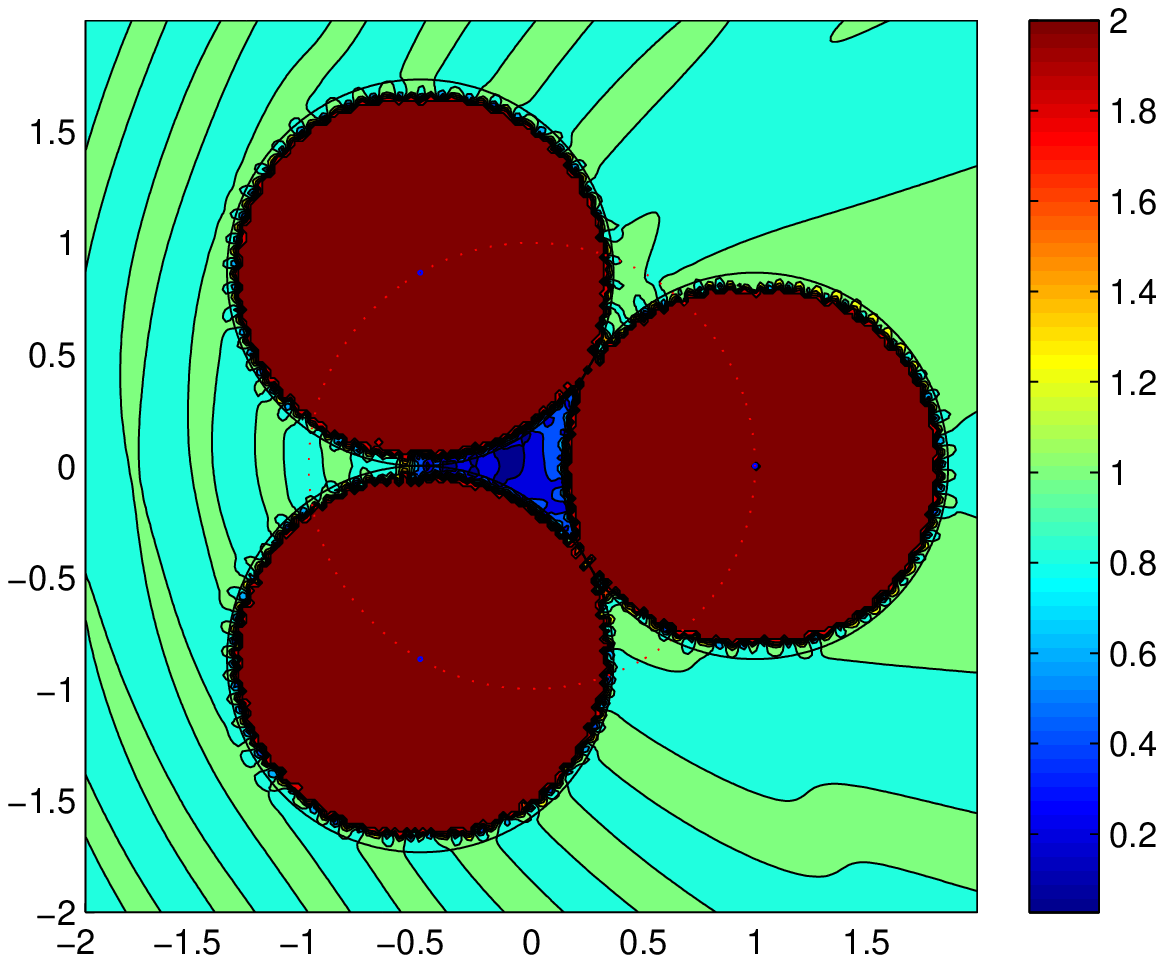}
   \label{fig:subfig66}
   }
   \caption[]{%
  Absolute  value of displacement vector components  $|u_y|/k_s$  for $N=5, k_p=2$ \subref{fig:subfig61}, $N=5, k_p=5$ \subref{fig:subfig62},  $N=20, k_p=2$ \subref{fig:subfig63},  $N=20, k_p=5$ \subref{fig:subfig64},   $N=50, k_p=2$ \subref{fig:subfig65}, and $N=50, k_p=5$ \subref{fig:subfig66} with $M = 3$ active sources for transverse wave incidence when cloaking devices are active.}
\label{fig:fig_60}
\end{figure}

The magnitude of the displacement vector  components  $u_x$ and $u_y$ are evaluated for  $\psi_p=7^\circ$ for various  values of  the truncation size $N$, the number of sources $M$, and the compressional wavenumber $k_p$. Greater accuracy is observed, as expected,   with increased $N$ and $M$.  However, large $N$ and $M$ require longer computation time, and some numerical experimentation is necessary to find the smallest values for which the displacement field vanishes to the desired degree in the cloaked region.

{
The magnitudes of ${|u_x|}/{k_p}$ and ${|u_y|}/{k_p}$   are depicted in Fig.\ \ref{fig:fig_20} and Fig.\ \ref{fig:fig_30}  for  longitudinal  incidence at different values of $N$  when cloaking devices are active with $M=3, k_p =2$. As expected,  the increase of $N$ is accompanied with the  reduction of magnitudes ${|u_x|}/{k_p}$ and ${|u_y|}/{k_p}$ in the cloaked region. 
}

{
Figure \ref{fig:fig_40} illustrates $|u_y|/k_p$ for  longitudinal  incidence with $N=5$ changing the values of $k_p$ and $M$ whilst Fig. \ref{fig:fig_50} and Fig. \ref{fig:fig_60} show corresponding values of $|u_x|/k_s$ and $|u_y|/k_s$ for  shear incidence, varying $N$ and $M$  with $k_p =2$ 
for the former, and  altering the values of $N$ and $k_p$ with $M=3$ for the latter.  Comparison of these results shows that at  higher frequencies, i.e., larger values of $k_p$, greater accuracy is achieved by increasing the number of sources $M$, whereas at lower frequencies the smallest  number of sources required, i.e. $M=3$, produces reasonable  cloaking, although this is enhanced with  increased values of  $N$.
}

\section{Conclusions}

The external active acoustic  cloaking model of \cite{Norris12b} has been  generalized to elastodynamics.  Just as in the former case  it is possible to represent the sources in exact terms, although it requires that the  incident elastic wave field is  known in exact form; this is  the price paid for active control.
The control method proposed is based on representing the incident field in terms of regular functions (Bessel functions) at each source position, which leads to a linear system of equations for the source amplitudes that can be solved in closed form.  The linear nature of the solution of  this essentially inverse problem means that arbitrary incident wave motion can be treated by superposition.

The results presented here provide a first step in the direction of realistic active control of elastic waves.  Applications to structure borne waves, surface waves, and even geophysical waves, are possible.  However, as a {\it{control}} problem, many issues remain  to be addressed.  Not least is the issue  of how to balance the goal of silencing one region of space with the unavoidable source noise that must be generated  in another, larger, region.  This quandary arises from the fact that the infinite series for the multipole expansion of the $m$ active source is divergent inside the domain $A_m$. Exact field cancellation is not achievable in practice; it becomes necessary to truncate the series and balance the decrease in  cloaking  accuracy  with whatever amplitude level is deemed  acceptable in the source region.   This is obviously a crucial  aspect and one  that remains to be studied in detail.  We have pointed out some similarities with parallel issues in active noise control, and future studies will examine analogies in these topics.  One area for consideration is the low frequency end of the spectrum.
  The numerical  simulations presented here indicate  that a small number of multipoles provide adequate cancellation at low frequencies. This suggests a natural way to extend ideas based on monopoles  to more elaborate sources distributions composed of finite numbers of multipoles of low order.  Hopefully, the present results provide a means to establish realistic strategies for practical application.


\section*{Acknowledgment}

This work was supported by NSF grant CMMI-0928499.



\end{document}